\documentclass[aps,prb,twocolumn,citeautoscript,nofootinbib,eqsecnum,10pt]{revtex4-1}  
\pdfoutput=1

\usepackage{amsmath}
\usepackage{amssymb}
\usepackage{bm} 
\usepackage{graphicx}
\usepackage[caption=false]{subfig}
\usepackage{color} 

\usepackage[breaklinks,colorlinks = true,linkcolor = red,urlcolor=red,citecolor=red]{hyperref} 
\hypersetup{
    bookmarks=true,         
    unicode=false,          
    pdftoolbar=true,        
    pdfmenubar=true,        
    pdffitwindow=false,     
    pdfstartview={FitH},    
    pdftitle={},    
    pdfauthor={Author},     
    pdfsubject={Subject},   
    pdfcreator={Creator},   
    pdfproducer={Producer}, 
    pdfkeywords={keyword1} {key2} {key3}, 
    pdfnewwindow=true,      
    colorlinks=true,       
    linkcolor=magenta, 
    citecolor=blue,        
    filecolor=magenta,      
    urlcolor=blue           
}

\usepackage{amsfonts}
\usepackage{upgreek}
\usepackage{slashed}
\usepackage{latexsym}
\usepackage{ulem}

\usepackage{wrapfig} 
\usepackage{scrpage2}						
\usepackage{wasysym}			
\usepackage{amsthm}
\usepackage{xcolor}
\usepackage{xy}
\usepackage{cancel} 
\usepackage{eucal} 
\usepackage{feynmp}    
\usepackage{mathrsfs}
\usepackage{soul}
\usepackage{wrapfig} 

\newcommand{\bal}{\begin{align}} 
\newcommand{\eal}{\end{align}}
\newcommand{\nn}{\nonumber \\}
\newcommand{\be}{\begin{equation}}
\newcommand{\ee}{\end{equation}}

\begin{document}
\title{Electric Field Response in Breathing Pyrochlores}

\author{Ipsita Mandal}
\affiliation{Laboratory of Atomic And Solid State Physics, Cornell University, Ithaca, NY 14853, USA}

\date{\today}

\begin{abstract}
We study the effects of a uniform electric field on the the ground state and excitations of the three-dimensional U(1) spin liquid phase of a breathing pyrochlore lattice, arising due to the coupling between the conventional (Maxwell) electric field and the emergent electrodynamics of the quantum spin ice material. This is an extension of the studies for isotropic pyrochlores in Phys. Rev. B 96, 125145 (2017) to the anisotropic case, as the lattice inversion symmetry is broken in breathing pyrochlores. The emergent photons are found to exhibit birefringence, analogous to the isotropic case. However, the system possesses a nonzero polarization even in the absence of an external electric field, unlike the isotropic pyrochlore. We also find that a sufficiently strong electric field triggers a quantum phase transition into new U(1) quantum spin liquid phases which trap $\pi$-fluxes of the emergent electric field. Such transitions are seen to occur even when the applied electric field is along a direction that does not show a phase transition in the isotropic limit.
\end{abstract}

\maketitle
\tableofcontents

\section{Introduction}

When interacting spins are placed on certain lattices causing geometric frustration, the energy cannot be minimized in a unique way. But nature obeys the laws of quantum mechanics, and spins do not point in a specific direction (quantum fluctuations) -- this helps the spins decide what they want to do. Such a frustrated system is actually governed by the laws of quantum mechanics at a macroscopic scale, and leads to the concept of spin liquids. The idea of a quantum spin liquid (QSL) as a paramagnetic ground state, without any long-range order (due to quantum fluctuations), was first proposed by Anderson \cite{anderson1,anderson2}.
The possible connection with high-T$_c$ superconductivity as the leading instability 
of a QSL \cite{anderson2} has
spurred interest in this concept to this date, with remarkable theoretical \cite{balents-nature,qsl-sondhi-roderich,qsl-review-lucile,qsl-review-roderich} and experimental \cite{qsl-expt1,qsl-expt2,qsl-expt3,qsl-expt4} advances.
QSLs have long-range entanglement and an intrinsic topological order, which are of
particular interest as potential platforms for quantum computation, intrinsically protected from  decoherence \cite{kitaev-anyons}.

The search for QSLs has revealed a large class of candidate rare-earth magnets on the pyrochlore lattice, which can potentially realize various version of three-dimensional (3D) U(1) QSL with possible gapless emergent photon and gapped spin-1/2 bosonic spinons, as well as bosonic magnetic monopoles -- the so called {\it quantum spin ice} (QSI) \cite{castelnovo2008magnetic,bramwell,gingras2014quantum}. The QSI phase is best thought of starting from the classical antiferromagnetic Ising model on the pyrochlore lattice made up of corner-sharing tetrahedra. Here, the classical ground states consist of two ``up" and two ``down" spins per tetrahedron, and thus constitute a macroscopically degenerate manifold, which is lifted by the quantum perturbations to give rise to the QSI state.

\begin{figure}[htb]
\includegraphics[width = 0.3 \textwidth]{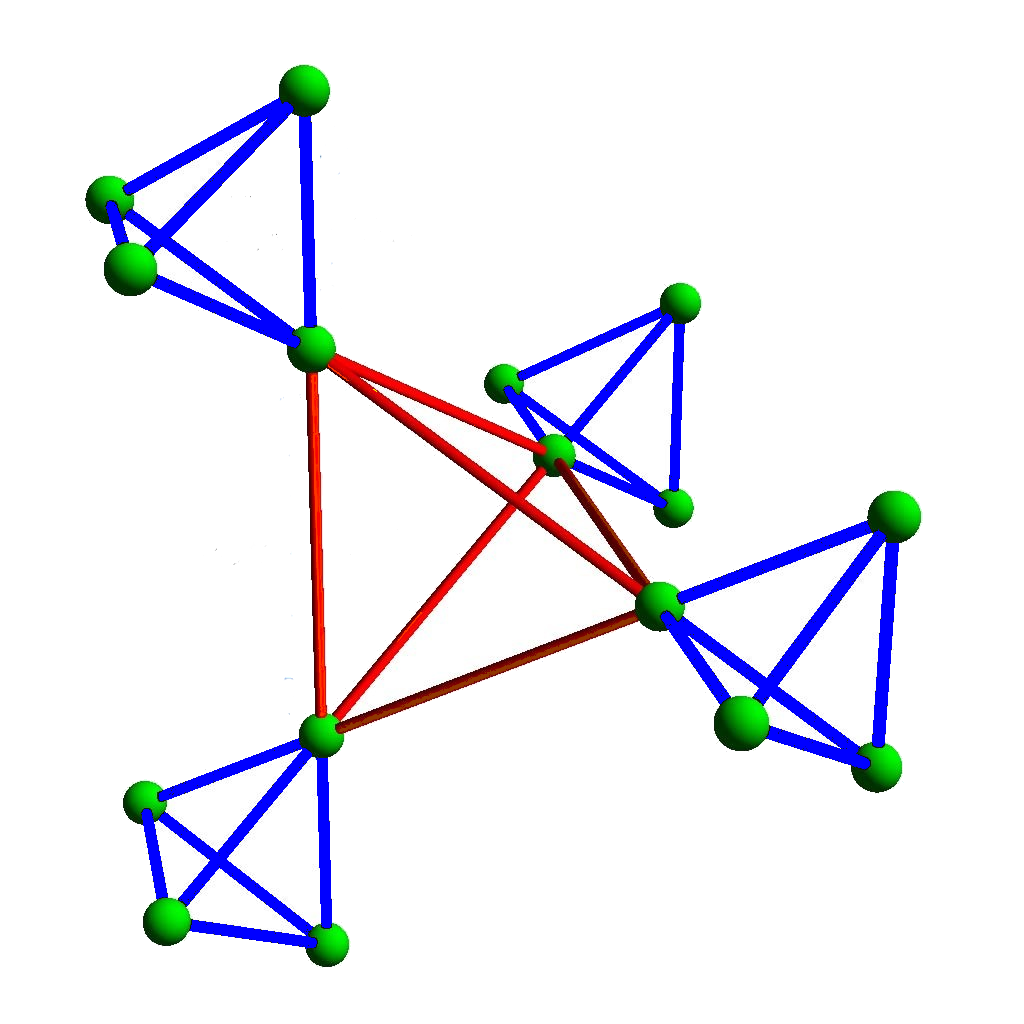}
\caption{The breathing pyrochlore lattice, whose bond lengths on the up-pointing and down-pointing tetrahedra are
different.}
\label{figtetra}
\end{figure}

In case of the rare-earth pyrochlores with chemical formula A$_2$B$_2$O$_7$ (where ``A" and ``B" are rare-earth or transition metal species), the spins of the A ions sit on the vertices of two kinds of tetrahedra -- the up-pointing and the down-pointing tetrahedra. These two types of tetrahedra are equivalent, as they are connected by inversion symmetry. Thus, the exchange couplings for the spins sharing a bond of the down-tetrahedron are equal to the same for the up-tetrahedron. Recently, the possibility of a {\it breathing pyrochlore} has been explored, where a staggered breathing inversion is present such that the lattice inversion symmetry is broken, as one set of the tetrahedra are smaller in size (see Fig.~\eqref{figtetra}). The spin exchanges are also accordingly scaled. The A-site spinels LiGaCr$_4$O$_8$ and LiInCr$_4$O$_8$ \cite{okamoto2013, tanaka2014}, as well as Ba$_3$Yb$_2$Zn$_5$O$_{11}$ \cite{rau2016,haku2016}, present examples of such breathing pyrochlore magnets. In particular, Ba$_3$Yb$_2$Zn$_5$O$_{11}$ presents a case with effective spin $S=1/2$ degrees of freedom on each Yb$^{3+}$ site, that determines the nature of the low energy magnetic state \cite{rau2016,haku2016}. This system remains magnetically disordered down to 0.38 K even while the Curie-Weiss temperature, $\Theta_{CW}$, is $\sim -7$ K \cite{kimura2014}. 

These observations have raised the possibility of realising QSL phases in breathing pyrochlores. The simplest of these QSLs is the {\it breathing} incarnation of the QSI phase, which itself is stable to the inversion symmetry breaking \cite{savary2016}. Of course in the extreme anisotropic limit, the tetrahedra gets decoupled leading to a trivial paramagnet. Thus, it is conceivable that this series of compounds can also access the very non-trivial and Landau-forbidden transition between a {\it topological} quantum paramagnet (the QSI) and a trivial paramagnet, as a function of the breathing anisotropy. In this paper, we explore the issue of experimental signatures of QSI phases, and possible phase transitions to decoupled paramagnets, in terms of their response to a static external electric field. The isotropic limit was studied in Ref.~\onlinecite{etienne}.

The paper is structured as follows. In Sec.~\ref{brpyr}, we outline the minimal Hamiltonian of the breathing pyrochlore lattice, and introduce the electric polarization operator consistent with the symmetries of the underlying lattice.
Sec.~\ref{secpert}, we first derive the low-energy effective theory of QSI in presence of a uniform external electric field, in the perturbative limit. Then we use this Hamiltonian to find the dispersions of the emergent photons of the U(1) QSL. We discuss the conditions when the system can undergo transitions to $\pi$-flux phases, signaled by the vanishing of the photon velocity, where certain hexagonal plaquettes of the tetrahedral structure trap a background (emergent) electric flux of $\pi$. We also chalk out the behavior of the spin structure factors, which can be observed in neutron scattering experiments. Finally, we compute the polarization of the material, finding an intrinsic nonzero polarization in the absence of any applied electric field.
We conclude with some discussions and outlook in Sec.~\ref{conclude}.
The appendices provide the details of some of the computations involved.

\section{Breathing Pyrochlores}
\label{brpyr}
\subsection{Pyrochlore Lattice}

Pyrochlore is a lattice of corner sharing tetrahedra with spin moments at the corners. For breathing pyrochlores, alternate tetrahedra are consistently expanded and contracted, as shown in Fig.~\eqref{figtetra}. For the four vertices of the tetrahedron, we use the same convention, as used in Ref.~\onlinecite{etienne}, for the local quantization axes. The centers of the tetrahedra form a diamond lattice, which comprises two intertwined FCC lattice structures. For the sake of completeness, we depict the convention in Appendix~\ref{appen_latticedetails}.

\subsection{Spin Hamiltonian}

\begin{figure}
\includegraphics[width = 0.5\textwidth]{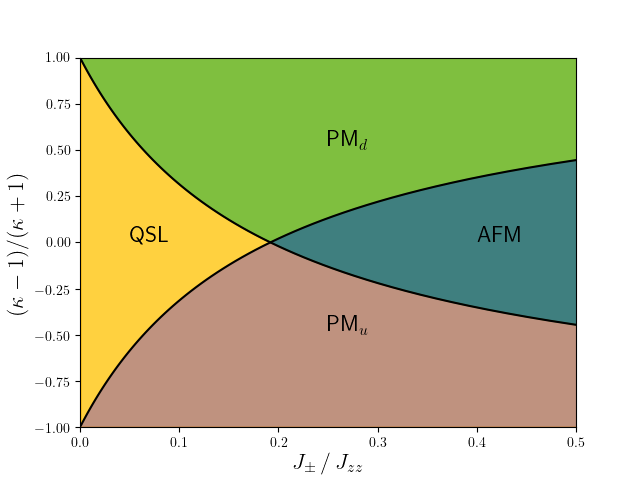}
\caption{Phase diagram of the minimal Hamiltonian of Eq.~\eqref{br-ham} for the breathing pyrochlores, in zero electric field. QSL and AFM denote the quantum spin liquid and antiferromagneic phases, respectively. The phases denoted by PM$_u$ and PM$_d$ are paramagnetic, consisting of decoupled up and down tetrahedra, respectively.}
\label{fig:phasediag_savary}
\end{figure}

The minimal Hamiltonian, consistent with symmetries, that describes the breathing pyrochlores, is given by \cite{savary2016}:
\begin{align}
H=&  J_{zz}\sum \limits_{\left< ij\right> \in u} s^z_i\,s^z_j
-J_\pm\sum_{\left< ij\right>}\left (s^+_i\,s^-_j+s^-_i\,s^+_j  \right ) \nonumber\\
& +  \kappa \left [  J_{zz}\sum \limits_{\left< ij\right> \in d }   s^z_i\, s^z_j
 -J_\pm\sum_{\left< ij\right>} \left (s^+_i\,s^-_j+s^-_i\, s^+_j \right ) \right ],
\label{br-ham}
\end{align}
where $J_{zz}>0$. For breathing pyrochlores, inversion  symmetry centered at lattice sites is absent as the bond lengths on an up-pointing tetrahedron (labeled by ``$u$")  and a down-pointing tetrahedron (labeled  by ``$d$") are unequal. This asymmetry of the crystal translates into an asymmetry of the exchange coupling for the spins as manifested in the above Hamiltonian.  The parameter $\kappa$ captures the asymmetry for $\kappa \neq 1$.

In Fig.~\eqref{fig:phasediag_savary}, we have reproduced the phase diagram obtained by Savary {\it et al} in Ref.~\onlinecite{savary2016}, using the gauge mean field theory (gMFT) machinery outlined in Appendix~\ref{appenpara}. Depending on  the sign of $\zeta = \frac{\kappa - 1}{\kappa +1}$, the $0$-flux QSL phase undergoes a transition to a paramagnetic phase. For $\zeta > 0$, the transition is to a paramagnet with decoupled up-tetrahedra; whereas for $\zeta <0$, a paramagnet with decoupled down-tetrahedra emerges.
For larger values of $\frac{J_\pm}{J_{zz}}$, there is a phase transition into an antiferromagnetic phase.

\subsection{Electric Polarization Operator}
While the bond-lengths of the up-tetrahedra and the down-tetrahedra are unequal, each of them still are undistorted tetrahedra, albeit of unequal sizes. Hence, at the single tetrahedron level, the symmetry analysis of the $T_d$ group alone can fix the form of the electric polarization operator. On carrying out this analysis, the electric polarization operator for an up-tetrahedron is given by \cite{etienne}:
\begin{align}
 \mathbf  P_u= {\bf P}^{(L)}+{\bf P}^{(T)}\,,
 \end{align}
where, as in Ref.~\onlinecite{etienne}, the longitudinal and the transverse parts are given by
 \begin{align}
{P}^{(L)}_x &= A \left (s_1^z\,s_4^z-s_2^z\,s_3^z  \right ),\nonumber\\
{P}^{(L)}_y &= A \left (s_1^z\,s_3^z-s_2^z\,s_4^z\right  ), \nonumber\\
{P}^{(L)}_z &= A \left (s_1^z\,s_2^z-s_3^z\, s_4^z \right  ),
\label{eq_plong}
\end{align}
and 
\begin{align}
{P}_x^{(T)}=& 
B\left[ \left (s^+_1\, s^-_4+\text{h.c.} \right )-\left (s^+_2\, s^-_3+\text{h.c.} \right  )\right],\nonumber \\
{P}_y^{(T)}=&
B\left[ \left (s^+_1\, s^-_3+\text{h.c.}\right )-\left (s^+_2\, s^-_4+\text{h.c.}\right )\right],\nonumber \\
{P}_z^{(T)}=& 
B\left[ \left (s^+_1\, s^-_2+\text{h.c.}\right )-\left (s^+_3\, s^-_4+\text{h.c.}\right )\right],
\label{eq_ptrans}
\end{align}
respectively. Here, $A$ and $B$ are unknown constants that depend on the microscopics.
Note that there may be other contributions to the transverse part. The most general form is summarized in Ref.~\onlinecite{etienne}. Given the electric polarization operator of an up-tetrahedron, the same for the down-tetrahedron is captured by:
\begin{align}
 \mathbf P_d = - \lambda\,  \mathbf P_u\,,
 \end{align}
where $\lambda >0\, (\neq 1)$ for breathing pyrochlores. 
The coupling to a uniform external electric field $\mathbf E$ is given by:
\begin{align}
H_{\bf E}= \sum_{\boxtimes\in {u} }{\bf P}_{{u}}\cdot{\bf E} +\sum_{\boxtimes\in {d} }{\bf P}_{{d}}\cdot{\bf E}\,,
\end{align}
Hence the full spin Hamiltonian, in the presence of an electric field, takes the form:
\begin{align}
\mathcal{H}=H+H_{\bf E}\,,
\label{breathing_haminfield}
\end{align}
where $H$ is given by Eq.~(\ref{br-ham}).

The longitudinal terms, which depend only on the $s^z$ terms, do not have an effect on the phase transitions that we wish to discuss. This is because all of them are identically zero (at the mean field level), and hereafter, we will drop the superscript ``$(T)$" from the transverse parts.

\section{Perturbation Theory within the QSI phase}
\label{secpert}

In the regime where $J_{zz}\gg J_\pm,\, |\mathbf E|$, the ground state of this system is a $U(1)$ QSL -- the QSI, with gapped electric and magnetic charges and gapless photons \cite{hermele}.
The low-energy description of the QSL phase can be obtained by starting with the Hamiltonian in Eq.~(\ref{breathing_haminfield}) and performing a perturbative expansion, extending the formalism in Ref.~\onlinecite{hermele,etienne}.

The contribution from the ring-exchange term is given by:
\begin{align}
\mathcal{H}^1_{ \text{eff} }=&-  \frac{ 6  J_\pm^3}{J_{zz}^2}(\kappa^3 +\frac{1}{\kappa^{2}}) \sum_{{\hexagon}}\left(\mathcal{O}^b_{\hexagon}+\text{h.c.}\right)\,,
\label{term1}
\end{align}
where $\mathcal{O}^b_{\hexagon}=s_1^+s_2^-s_3^+s_4^-s_5^+s_6^-  
$ ($1,\dots 6\in {\hexagon}$) is an operator that flips a loop of spins on hexagons. 
The $\kappa^3$ factor comes when we choose to flip the spins of the up-tetrahedra of the hexagon. Similarly, the $\kappa^{-2}$ originates from spin-flips on the participating down-tetrahedra. 

The new contribution, which is particular to the anisotropic case, appears at $\mathcal O ( B \, J_{\pm}^2)$, and is given by:
\begin{align}
\label{term2}
\mathcal{H}^2_{ \text{eff} }=& \frac{6\, \sqrt 3 \left[\lambda\, \kappa^2  - \frac{1}{\kappa^{2} } \right]   B \,J_\pm^2} {J^2_{zz}}  \sum \limits_{  \hat{\mathbf{t}}_m=1} ^4   \left[\mathbf E \cdot \hat{\mathbf{t}}_m \right]\nonumber\\
&~~~~~~~~~~~~~~~~~~~~~~~~~~~~~~~~~~\times\sum_{{\hexagon}\perp   \hat{\mathbf{t}}_m} \left[\mathcal{O}^b_{\hexagon}+\text{h.c.}\right] ,
\end{align}
with the $  \hat{\mathbf{t}}_m $'s defined in Eq.~(\ref{eq_t_axes}).
For example, for the hexagon which is perpendicular to $\hat {\mathbf t}_3$, we have a term proportional to $\left (  \lambda\, \kappa^2-\frac{1}{\kappa^2}\right ) \left (E_x-E_y+E_z
\right )$. Clearly this term vanishes in the isotropic limit when $\kappa=\lambda=1$. This is expected as this term is linear in the external electric field, which is odd under inversion. Such a linear term cannot survive in the perturbation theory around an inversion symmetric state, which is the case in the isotropic limit.

There is a contribution at $\mathcal O ( B^3 )$, given by:
\begin{align}
\label{term3}
\mathcal{H}^3_{ \text{eff} }=& \frac{ 6  \left ( \lambda^3 -\frac{1}{\kappa^2} \right )  B^3 E_xE_yE_z} { J^2_{zz}}  
\sum_{{\hexagon}} \left(\mathcal{O}^b_{\hexagon}+\text{h.c.}\right) .
\end{align}

Let us define the variables:
\begin{align}
g = \frac{12 \,J_\pm^3}{J_{zz}^2}\,, 
 \quad \gamma = \frac{B \,E}{J_\pm} \,.
\end{align}
\begin{figure}[htb]
\subfloat[]{
\includegraphics[width = 0.45\textwidth]{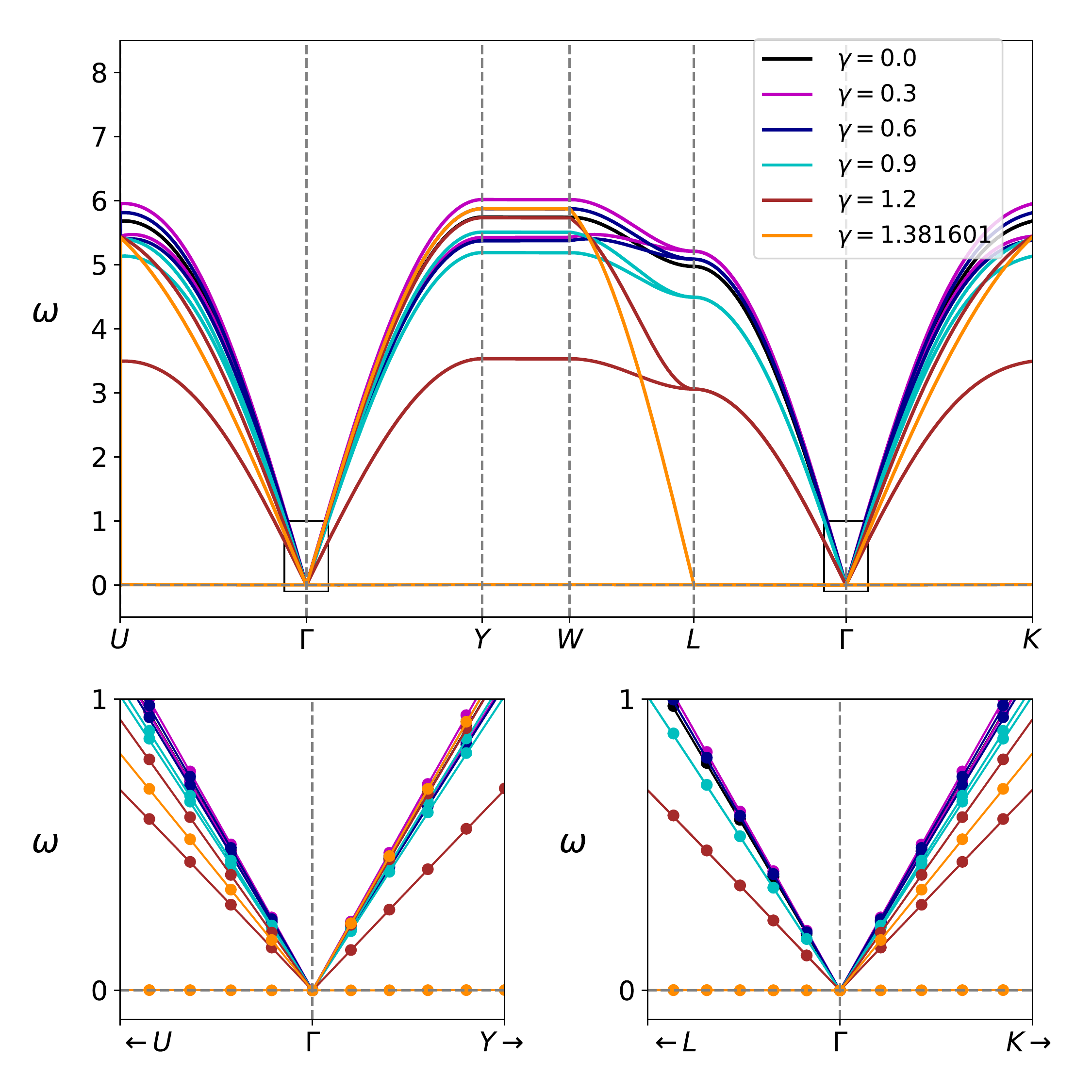}
}\quad
\subfloat[]{
\includegraphics[width = 0.45\textwidth]{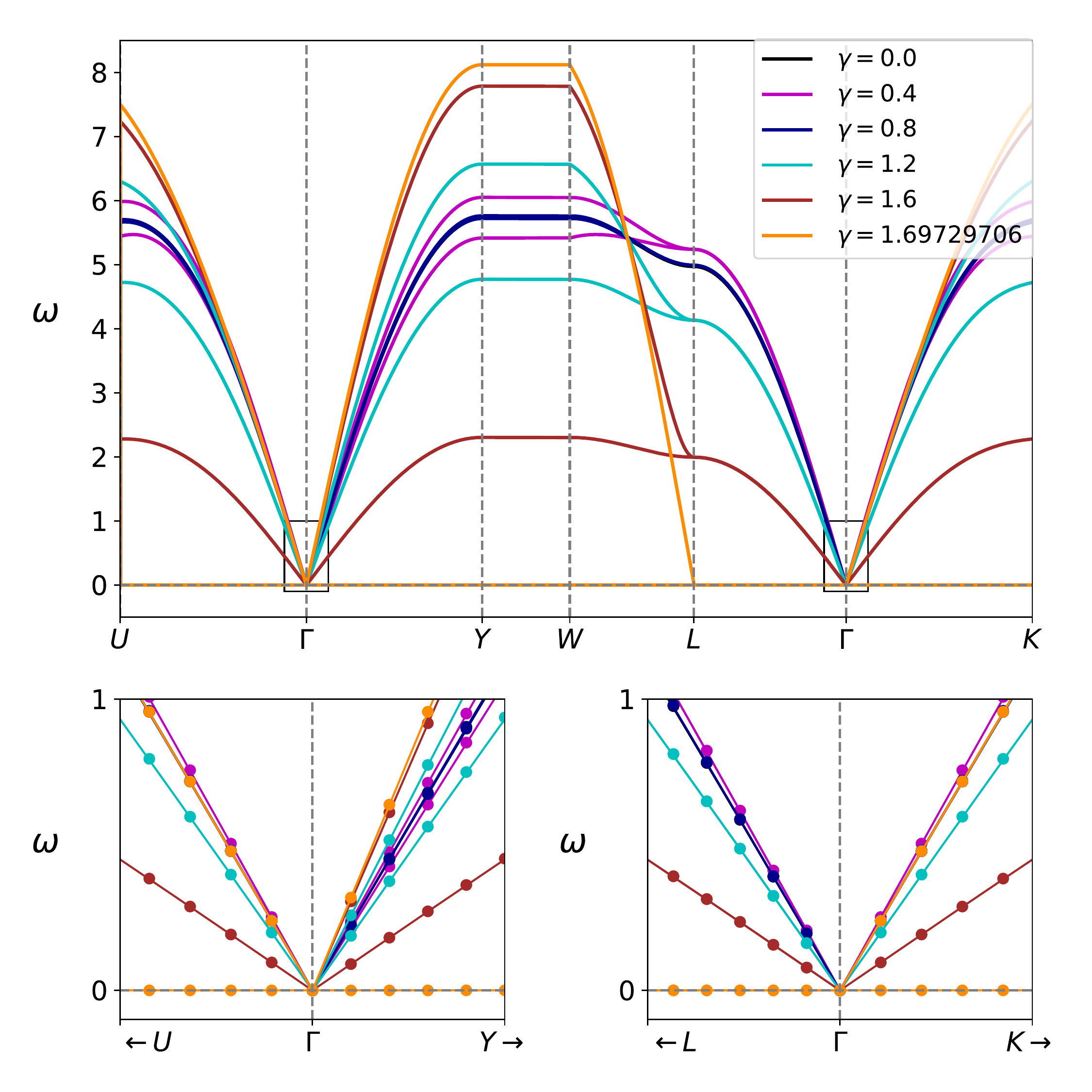}
}
\caption{Comparison of the spectra of the emergent photons for ${\bf E}  \propto  [111]$ and $\kappa=\lambda=0.5$, where the $\mathcal{H}^3_{eff}$ contribution is (a) included, and (b) dropped.
The lower panels show the results for the continuum limit discussed in Sec.~\ref{seccont}.}
\label{fig:compare}
\end{figure}
The terms in $\mathcal{H}^3_{ \text{eff} }$, along with $ \mathcal{H}^1_{ \text{eff} }$, are the isotropic contributions. Collecting both of them gives a renormalised $g'$ of Eq.~\eqref{term1}, captured by:
\begin{align}
g'=&g\left[\kappa^3 + \frac{1}{\kappa^2}-\left ( \lambda^3 -\frac{1}{\kappa^2} \right )\gamma^3
\cos\theta_{ex} \cos\theta_{ey} \cos\theta_{ez}\right] ,
\end{align}
where $\cos\theta_{ex}, \,\cos\theta_{ey}$ and $\cos\theta_{ez}$ denote the angles between $\mathbf E$ and the
$x,\,y$ and $z$ axes, respectively. 
While this term renormalizes the velocity isotropically (and hence changes the phase boundaries and other such details), it does not cause any other qualitatively new effect. 
Plots of the spectra for $\mathbf E\propto [111]$ and $\kappa=\lambda = 0.5$, in Fig.~\eqref{fig:compare}, support this fact. For this reason, we will drop this term for simplicity.

The contribution at $\mathcal O ( B^2   J_{\pm} )$ is given by:
\begin{align}
\label{term4}
\mathcal{H}^4_{ \text{eff} }=& \frac{6 \left (  \lambda^2 \,{\kappa } + \kappa^{-2} \right )  B^2 J_\pm } { J^2_{zz}}  \sum \limits_{  \hat{\mathbf{t}}_m=1} ^4
\left [  \mathbf E \cdot \mathbf E -3 \left ( \mathbf E \cdot \hat{\mathbf{t}}_m \right )^2 \right  ]\nonumber\\
&~~~~~~~~~~~~~~~~~~~~~~~~~~~~\times \sum_{{\hexagon}\perp   \hat{\mathbf{t}}_m} \left(\mathcal{O}^b_{\hexagon}+\text{h.c.}\right) .
\end{align}

Collecting everything, the total effective Hamiltonian is then given by:
\begin{align}
\label{termtot}
\mathcal{H}_{ \text{eff} }=&  \mathcal{H}^1_{ \text{eff} }  + \mathcal{H}^2_{ \text{eff} } + \mathcal{H}^4_{ \text{eff} } \,.
\end{align}

\subsection{Effective Low-energy Hamiltonian}

\begin{figure*}[htb]
\subfloat[]{
\includegraphics[width =  0.3 \textwidth]{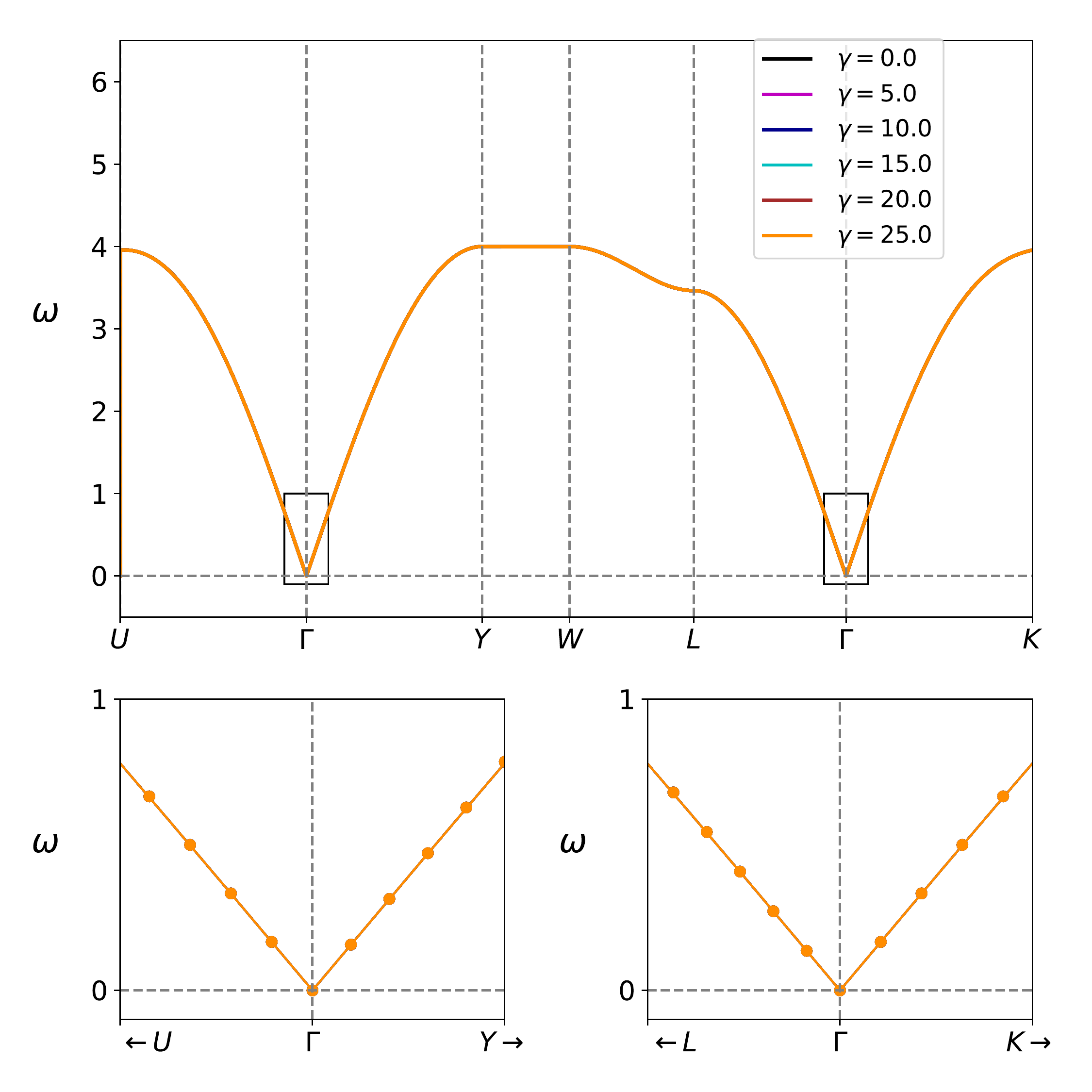}
}\quad
\subfloat[]{
\includegraphics[width =  0.3 \textwidth]{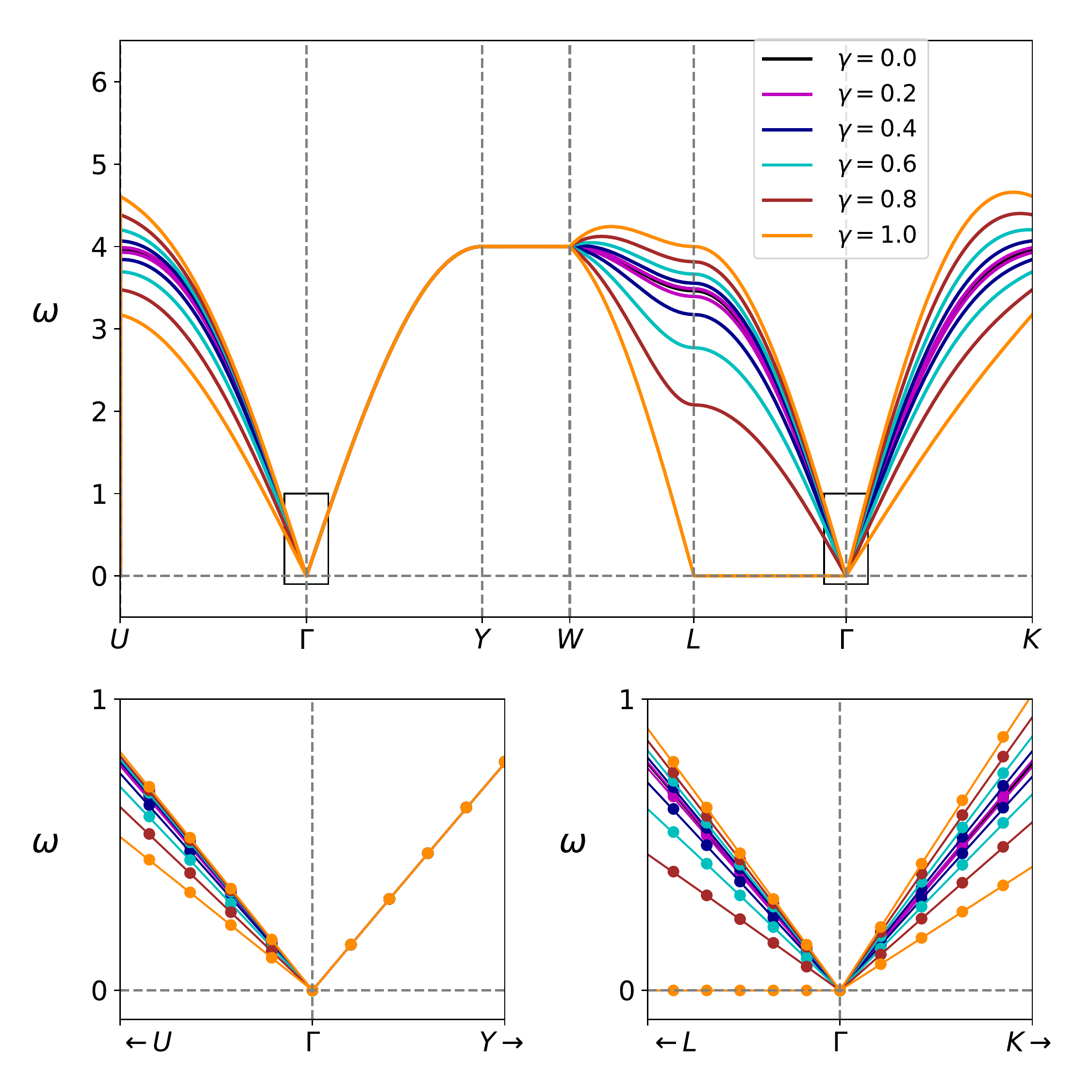}
}\quad
\subfloat[]{
\includegraphics[width =  0.3 \textwidth]{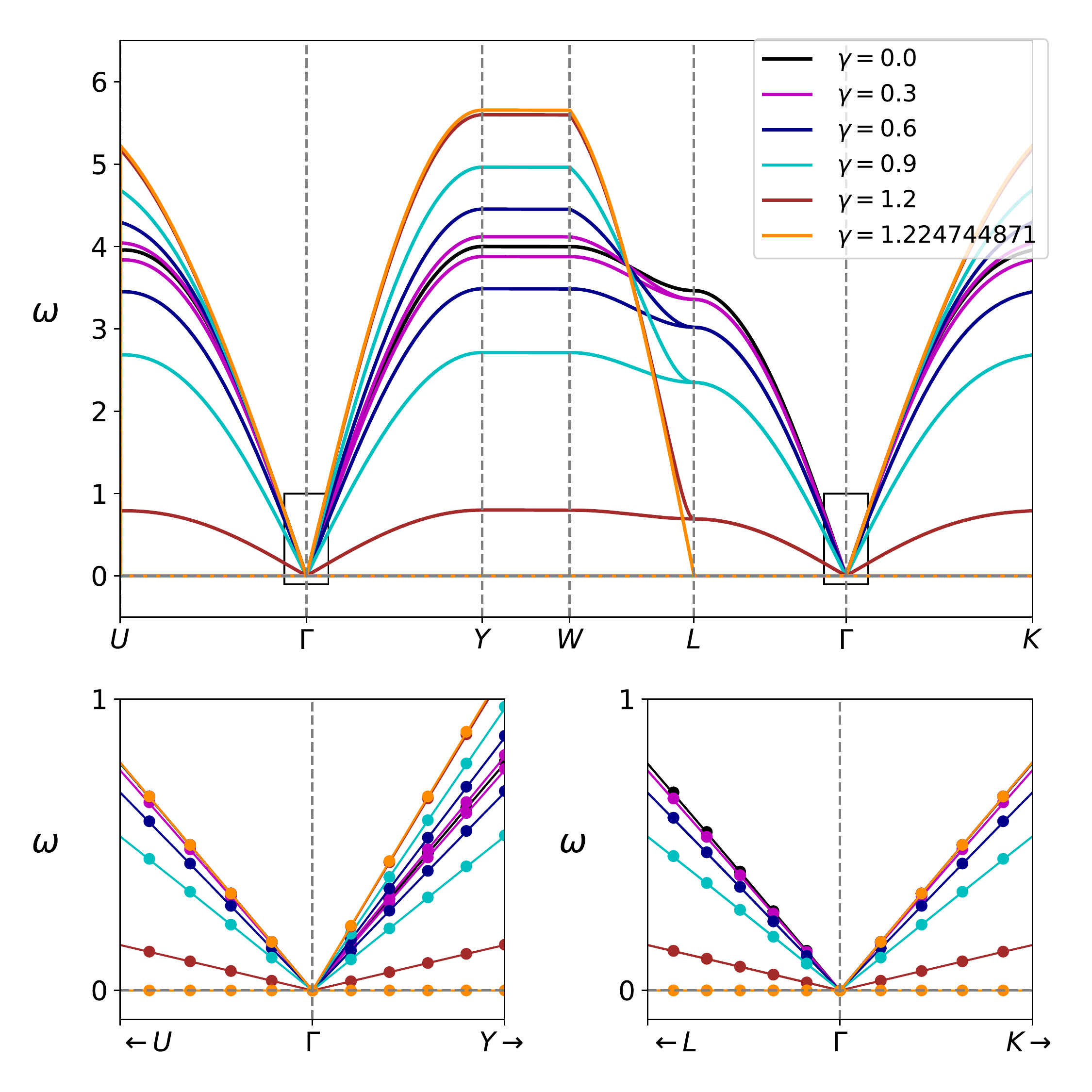}
}
\caption{Spectra of the emergent photons for the isotropic pyrochlore case, when (a) ${\bf E} \propto [010]$ (b) ${\bf E} \propto  [011]$, and (c) ${\bf E} \propto  [111]$, with $\kappa =0.5$.}
\label{fig:freq1}
\subfloat[]{
\includegraphics[width = 0.3 \textwidth]{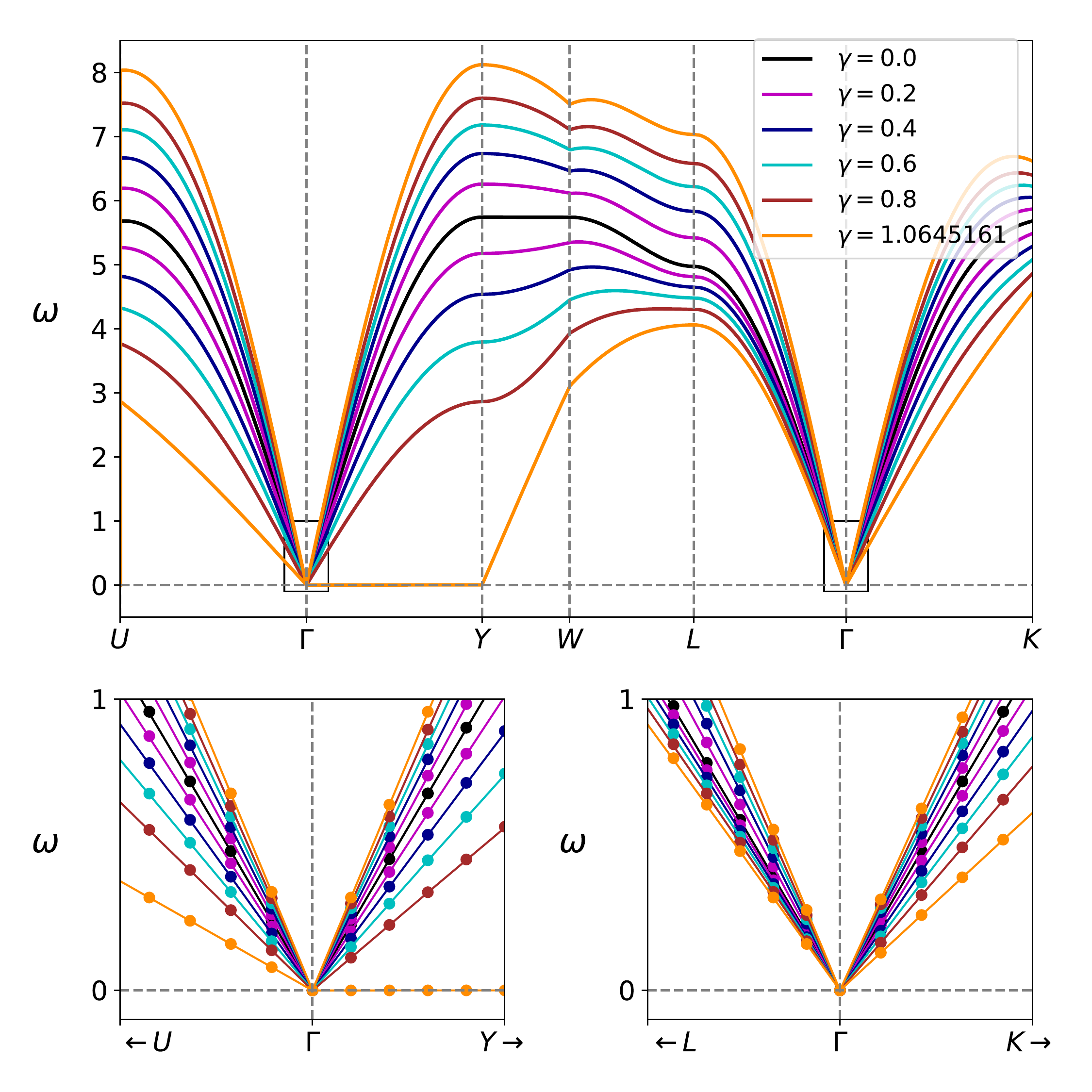}
} \quad
\subfloat[]{
\includegraphics[width = 0.3\textwidth]{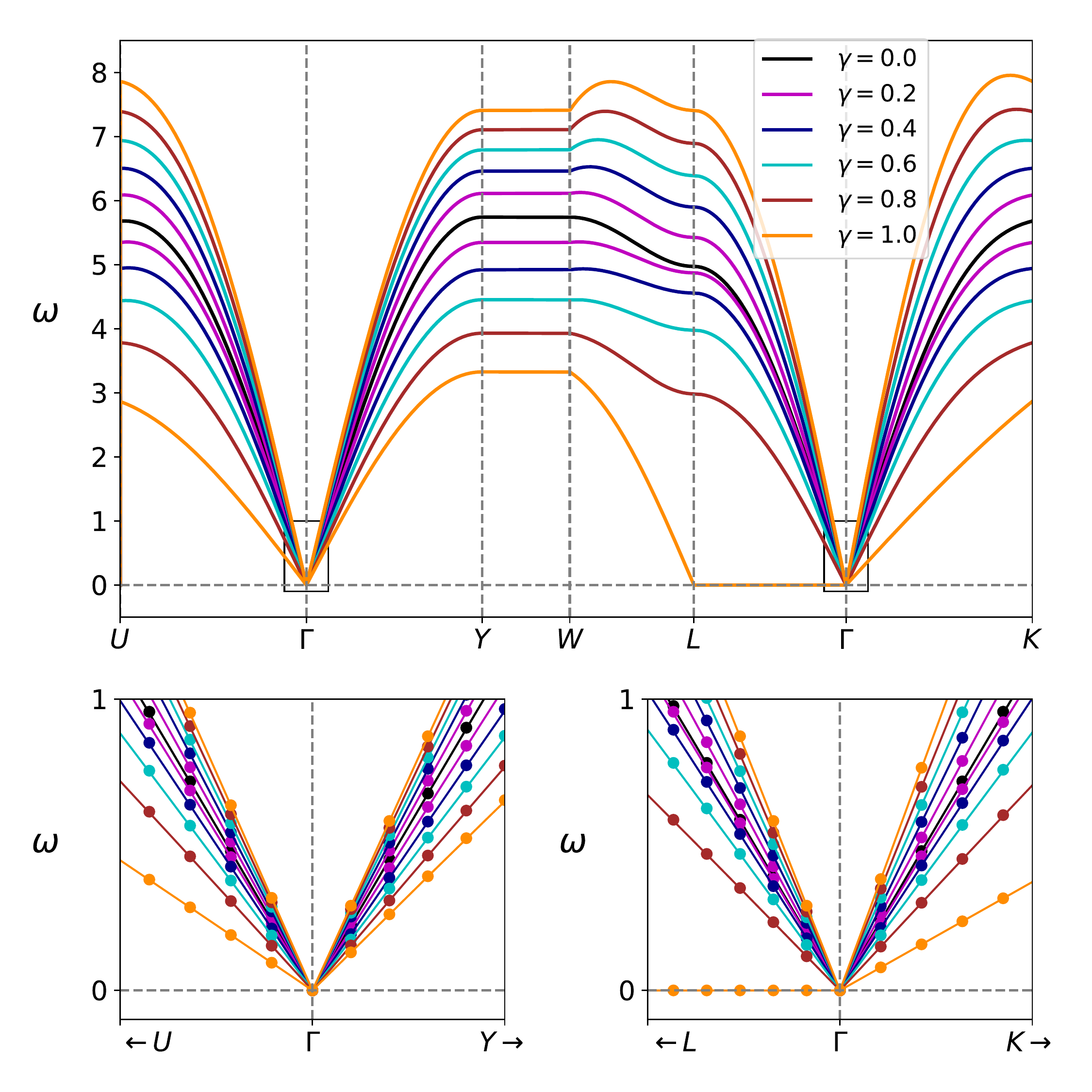}
} \quad
\subfloat[]{
\includegraphics[width = 0.3\textwidth]{Spectrum_direction1_kappa0-5_lambda0-5_noheff3}
}
\caption{Spectra of the emergent photons along three directions of the electric field for the breathing pyrochlore case, when (a) ${\bf E} \propto [010]$, (b) ${\bf E} \propto [011]$, and (c) ${\bf E} \propto [111]$, with $\kappa = \lambda = 0.5$. }
\label{fig:freq2}
\end{figure*}

The low-energy physics of the system, encoded within
the effective Hamiltonian in Eq.~\eqref{termtot}, can be seen explicitly by mapping it to an effective problem of
electromagnetism \cite{hermele}, which captures the U(1) QSL picture. We employ the same notation and steps
as in Ref.~\onlinecite{etienne}, which we briefly describe below.
The spins sit on the
bonds of the diamond lattice that is obtained
by joining the centers of the tetrahedra forming the pyrochlore lattice. The site $i$ of the pyrochlore lattice can be thought of as the midpoint of the bond 
$\mathbf{r} \to \mathbf{r}'$ of a diamond lattice. 
Since this diamond lattice is bipartite, it is possible to define directed variables 
on these bonds as:
\begin{eqnarray}
b_{\mathbf{r} \mathbf{r}'}
    = -b_{\mathbf{r}' \mathbf{r}}\,, \quad \alpha_{\mathbf{r} \mathbf{r}'}
    = -\alpha_{\mathbf{r}' \mathbf{r}} \,,
\end{eqnarray}
through the mapping
\begin{eqnarray}
\label{eq:ab}
b_{\mathbf{r} \mathbf{r}'} &=& \pm \left( \hat{n}_i-\frac{1}{2} \right) \,,\quad
\alpha_{\mathbf{r} \mathbf{r}'} = \pm \theta_i\,,
\end{eqnarray}
where the sign is taken to be positive if $\mathbf{r}$ belongs to the 
$A$-sublattice, and negative if $\mathbf{r}$ belongs to the $B$-sublattice.   
Taking this convention into account, we are left with a pair
of canonically conjugate variables:
\begin{eqnarray}
\label{BGcanon}
\left[ \alpha _{\mathbf{r} \mathbf{r}'}, b_{\mathbf{r}'' \mathbf{r}'''}\right] 
   = i \left(\delta_{\mathbf{r} \mathbf{r}''} \delta_{\mathbf{r}' \mathbf{r}'''} 
    -\delta_{\mathbf{r} \mathbf{r}'''} \delta_{\mathbf{r}' \mathbf{r}''}  \right).
\end{eqnarray}
The field $b_{\mathbf{r} \mathbf{r}'}$ plays the role of an emergent magnetic field.  
The corresponding electric field, $ e_{\mathbf{s} \mathbf{s}'}$, inhabits
the bonds $\mathbf{s} \to \mathbf{s}'$ of a second diamond lattice, 
interpenetrating the first, and is defined through a lattice curl: 
\begin{eqnarray}
e_{\mathbf{s} \mathbf{s}'}
= (\nabla_{\scriptsize\hexagon} \times \alpha )_{\mathbf{s} \mathbf{s}'} 
= \sum_{\circlearrowleft} \alpha_{\mathbf{r} \mathbf{r}'} \,,
\label{eq:Ecurl}
\end{eqnarray}
where the sum $\sum_{\circlearrowleft}$ is taken in an {\it anticlockwise} sense,
around the hexagonal plaquette of pyrochlore lattice sites encircling the bond 
$\mathbf{s} \to \mathbf{s}'$.   
Clearly, \begin{eqnarray}
e_{\mathbf{s} \mathbf{s}'}= - e_{\mathbf{s}' \mathbf{s}} \,.
\end{eqnarray}
We now introduce a complementary labeling convention which will be convenient for our calculations.  We denote the bonds on the direct and dual diamond lattices, respectively, by $({\bf r},n) \equiv (\mathbf{r}, \mathbf{r} + \mathbf{t}_n)$ and $({\bf s},m) \equiv (\mathbf{s}, \mathbf{s} + \mathbf{t}_m$), where $\mathbf{t}_n = \frac{a_0 \sqrt{3}}{4} \hat{\mathbf{t}}_n$. These represent bonds connecting the up-tetrahedron center $\mathbf{r}$ ($\mathbf{s}$) to the closest down-tetrahedron center in the direction of the spin quantization axis $\hat{\mathbf{t}}_n$ ($\hat{\mathbf{t}}_m$). Using this notation, finally, we can rewrite Eq.~(\ref{termtot}) as:
\begin{align}
\mathcal{H}_{\text{eff}}
 =&\frac{U}{2}\sum_{ {\bf r}, n} b_{ \mathbf{r}, n}^2 - \sum_{\mathbf{s},m} \mathcal{M}_m
 \cos \left( e_{\mathbf{s}, m} \right ), \label{eq_lattice_ham_matrixform} 
\end{align}
where $U$ is a model parameter, and
\begin{align}
\label{eq_Mmm}
\mathcal{M}_m  & =  g \Big [ \kappa^3 +\kappa^{-2}  +
 \gamma^2   \left (  \lambda^2 \,{\kappa } +  \kappa^{-2}   \right ) 
 \left \lbrace 3  \left ( \hat {\mathbf E} \cdot \hat{\mathbf{t}}_m \right )^2  -1 \right \rbrace 
 \Big .\nonumber\\
& ~~~~~~~~~~~ \left.-    \sqrt{3\,}\gamma \left (  { \kappa^{-2}}  -  \lambda\, \kappa^2 \right  ) \left (\hat{ \mathbf E} \cdot \hat{\mathbf{t}}_m \right )\right].
\end{align}


\begin{figure*}[htb]
\subfloat[]{
\includegraphics[width = 6cm]{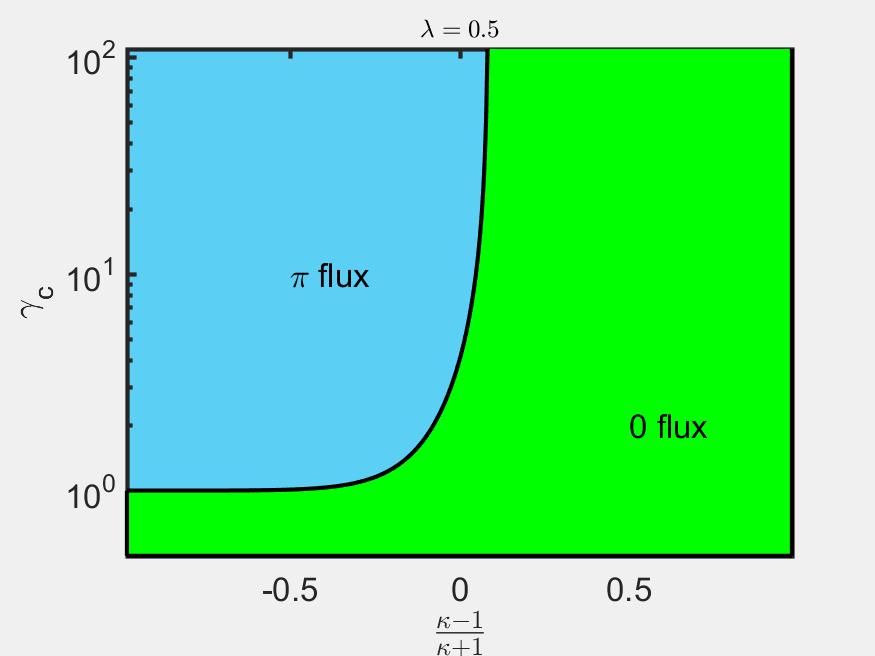}
}
\subfloat[]{
\includegraphics[width = 6cm]{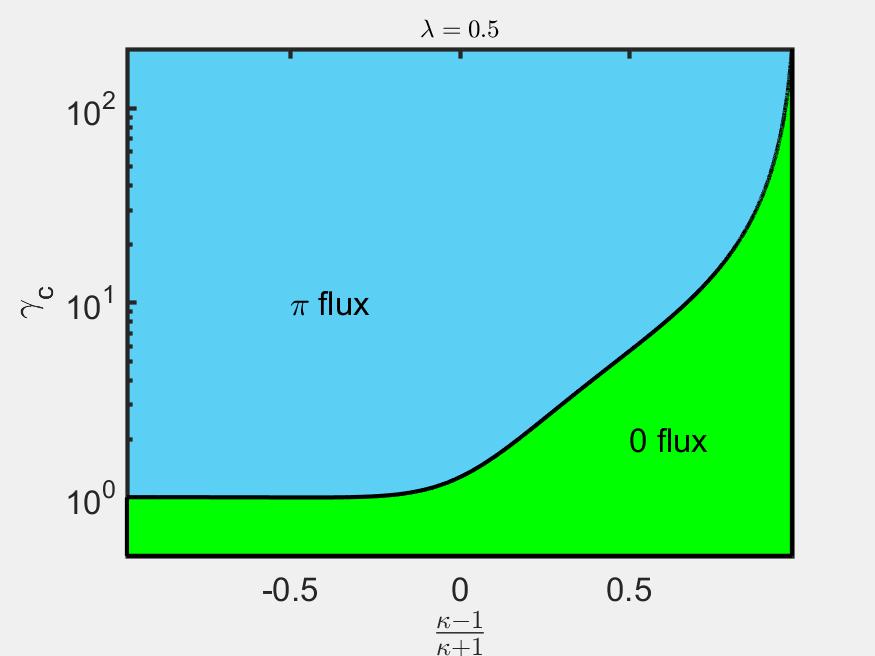}
}
\subfloat[]{
\includegraphics[width = 6cm]{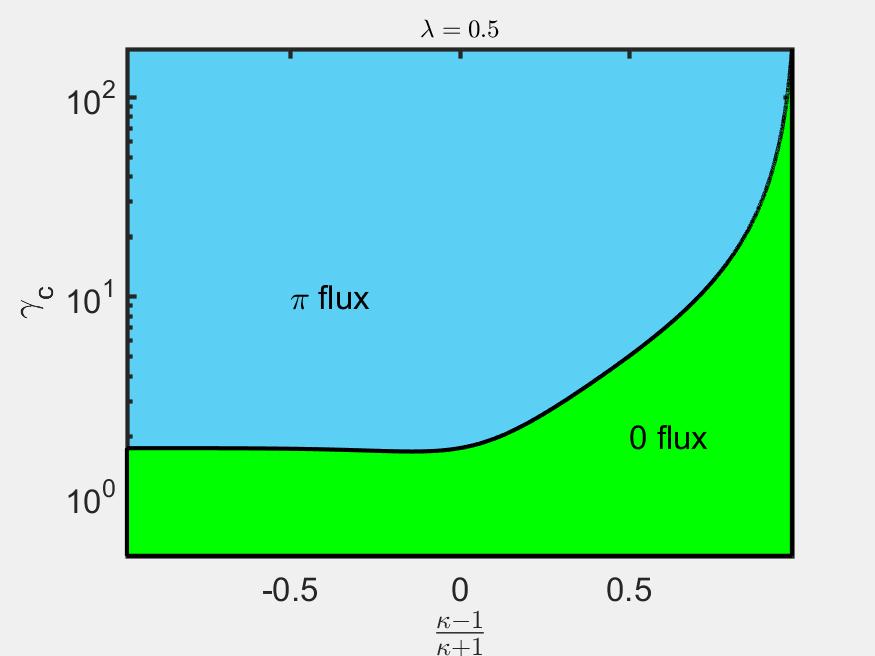}
}
\caption{Phase diagram showing the $0$-flux and the $\pi$-flux phases for the breathing pyrochlore, with $\lambda = 0.5$, and (a) $\mathbf E \propto  [010]$; (b) $\mathbf E \propto [011]$; (c) $\mathbf E \propto [111]$.}
\label{fig:phasediag_noflux-flux}
\end{figure*}

\subsection{Continuum Limit}
\label{seccont}

We first look at the continuum limit of the system before performing the lattice calculations. In this limit, the effective continuum Hamiltonian takes the form:
\begin{align}
\mathcal{H}_\text{eff}^\text{cont} &=\sum_\phi \int \frac{d^3\mathbf{k}}{(2\,\pi)^3}  \omega_\phi(\mathbf{k}) \Big[a_{\phi \,\mathbf{k}}^\dagger\, a_{\phi\, \mathbf{k}} + \frac{1}{2} \Big]\,.
\label{cont-ham}
\end{align}
The details of this derivation can be found in Appendix~\ref{continuum}.

The limit $\chi = \gamma^2 < 1 $ corresponds to the small field limit after we fix $2\, U g = 1$. It is important to note that this continuum theory is valid only for long wavelengths, i.e., near the center of the Brillouin zone.  In the continuum limit , the comparison of results for the isotropic and the breathing pyrochlores are shown in the lower panels of Fig.~\eqref{fig:freq1} and \eqref{fig:freq2}, for different directions of the electric field.

\subsection{Photon Dispersions in the Lattice Theory}

In order to understand the dispersion of the emergent photons throughout the Brillouin zone, we now perform the full lattice calculations, following Ref.~\onlinecite{benton,etienne} (replacing their $\mathcal{M}_m $ with the
one defined in Eq.~\eqref{eq_Mmm}). The diagonalization of the Hamiltonian is carried out in the same way as in Ref.~\onlinecite{etienne}. The results are shown in Fig.~\eqref{fig:freq2}, when $\kappa=\lambda=0.5$. To compare these results with the isotropic pyrochlore case of Ref.~\onlinecite{etienne}, we also show Fig.~\eqref{fig:freq1}.

For the $\mathbf E\propto [011]$ case, we note that for the isotropic limit, the dispersion along the $\Gamma-Y-W$ high symmetry direction is independent of $\gamma$, and both the polarization modes behave in an identical way. This symmetry is lifted for the breathing pyrochlore case.

For $\mathbf E\propto [111]$, along the $L-\Gamma$ high symmetry direction, the two polarization modes become degenerate for the isotropic case, a feature that remains intact even for the breathing pyrochlores.

\subsection{$\pi$-flux Phases}
\label{piflux}

In absence of an external electric field, the isotropic system described by Eq.~\eqref{term1} with $\kappa=1$, has $0$ emergent electric flux through each hexagonal plaquette (for $g>0$). For $\kappa\neq 1$, this remains valid, as the sign of the coupling constant does not change sign (unless the magnetic interactions on the up and down tetrahedra are opposite).
However, this is no longer true in presence of an external electric field $\mathbf E$, both for the isotropic as well as the breathing cases.
As the strength of the external electric field is increased, the
photon velocity vanishes at a critical value $\gamma =\gamma_c$, signaling an instability of the
low-field $0$-flux QSL state to a $\pi$-flux phase of the emergent electric field \cite{chang2012higgs,gang-chen}. 
As discussed in Ref.~\onlinecite{etienne}, the condition for trapping a $\pi$-flux through a hexagonal plaquette,
perpendicular to $\hat{\bf t}_m$,
corresponds to a change of the sign of $\mathcal{M}_m$ (defined in Eq.~\eqref{eq_Mmm}).
Figs.~\eqref{fig:freq1}  and~\eqref{fig:freq2} shows the different scenarios for some representative parameter values for the isotropic and anisotropic cases, respectively. 

\begin{figure*}
\subfloat[]{	
	\includegraphics[width=0.48\textwidth]{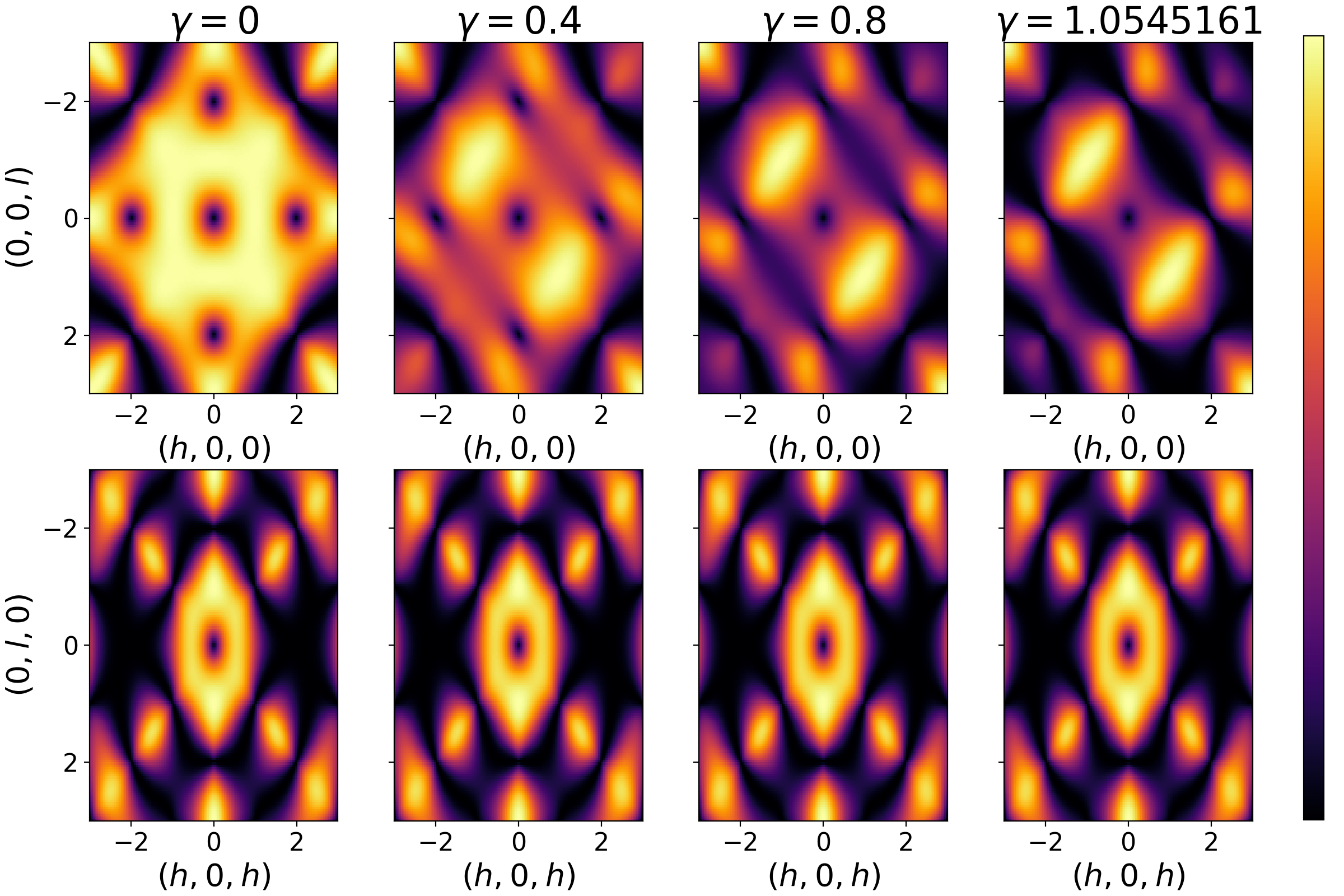}}
	\quad
\subfloat[]{
\includegraphics[width=0.48\textwidth]{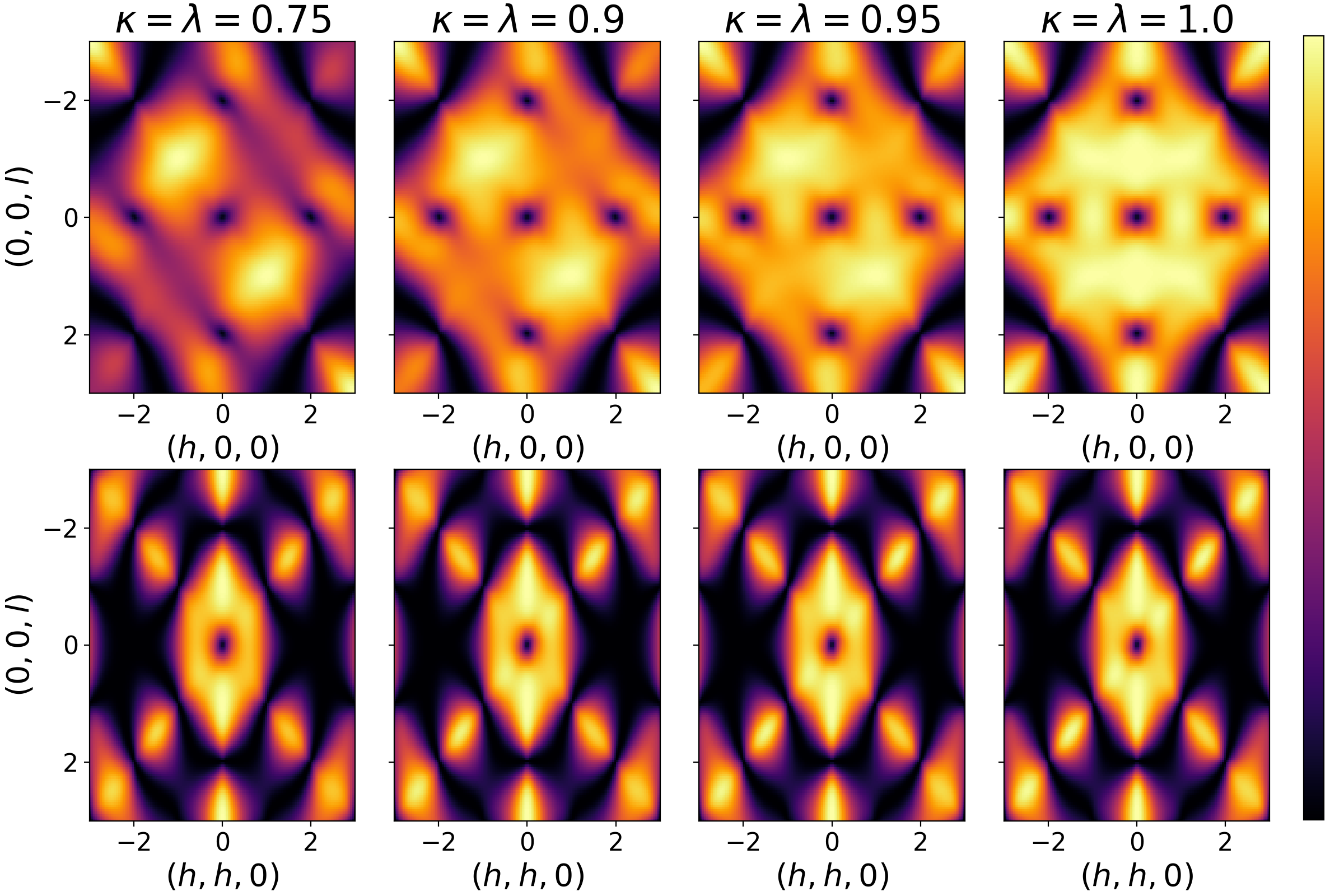}}
\caption{Equal-time spin structure factors (see Eq.~(\ref{eq_structurefactors})) in the spin-flip channel.
We have used $\frac{ \mathcal{W} J_{zz}^2} { U J_\pm^3} = 0.1$ to regularize the theory near the phase transitions, as explained in Ref.~\onlinecite{etienne}. Each subplot has an independent color scale. 
(a) Here, ${\bf E} \propto [010]$ and $\kappa = \lambda = 0.5$. The top panels show the $[h0l]$ scattering plane (using polarized neutrons with $\bm{\eta}_\nu \propto [010]$), and the bottom panels show the $[hlh]$ scattering plane (using polarized neutrons with $\bm{\eta}_\nu \propto [10\overline{1}]$). 
(b) Here ${\bf E} \propto [011]$, $\gamma = 1$, and different values of $\kappa=\lambda$ have been used. The top panels show the $[h0l]$ scattering plane (using polarized neutrons with $\bm{\eta}_\nu \propto [010]$), and the bottom panels show the $[hhl]$ scattering plane (using polarized neutrons with $\bm{\eta}_\nu \propto [1\overline{1}0]$).}
\label{fig:struct_fac}
\end{figure*}
 
For the isotropic case ($\kappa=\lambda=1$), this is given by
\begin{align}
1-3 \left (\hat{\bf E}\cdot\hat{t}_m \right )^2>1/ \gamma^2 \,.
\label{eq_condition_flux}
\end{align}
The $ {\bf E}\propto [010]$ case does not have a $\pi$-flux phase, as the above condition is never satisfied. For $\hat{\mathbf{E}} =\frac{ 1} {\sqrt{2}} \left( 0,1,1\right )$, we have
\begin{align*}
\left (\hat{\mathbf{E}} \cdot \hat{\mathbf{t}}_{1(4)}  \right)^2 =  2/3\,, \quad \left (\hat{\mathbf{E}} \cdot \hat{\mathbf{t}}_{2(3)}   \right )^2 = 0\,. 
\end{align*} 
Using Eq. (\ref{eq_condition_flux}), we see that the hexagons oriented perpendicular to $\hat{\bf t}_1$ and $\hat {\bf t}_4$ never trap a flux, while those oriented perpendicular to $\hat{\bf t}_2$ and $\hat {\bf t}_3$ do so  for $  \gamma^2 > 1$. In the dual lattice, they correspond to one-dimensional strings of electric flux running along $[0\bar{1}1]$ directions.
For $\hat{\mathbf{E}} = \frac{1}{\sqrt{3}} \left( 1,1,1 \right)$, we have 
\begin{align*}
\left (\hat{\mathbf{E}} \cdot \hat{\mathbf{t}}_1 \right )^2 = 1 \,, \quad \left(\hat{\mathbf{E}} \cdot \hat{\mathbf{t}}_{2(3,4)}\right )^2  = \frac{1}{9}\,.
\end{align*}
Using Eq.~(\ref{eq_condition_flux}), we see that the hexagons oriented along $\hat{\bf t}_1$ never trap a flux, but those oriented along  $\hat{\bf t}_2, \hat{\bf t}_3$ and $\hat{\bf t}_4$ trap fluxes for $ \gamma^2 > 3/2$. 

For the anisotropic case, the condition for trapping a flux through a hexagon (whose perpendicular is $\hat{\bf t}_m$) is given by
\begin{align}
& \Big [ \kappa^3 +\kappa^{-2}  +
 \chi   \left (  \lambda^2 \,{\kappa } +  \kappa^{-2}   \right ) 
 \big \lbrace 3  \left ( \hat {\mathbf E} \cdot \hat{\mathbf{t}}_m \right )^2  -1 \big \rbrace 
 \Big .\nonumber\\
& ~~~~~~~~~~~ \left.-    \sqrt{3\,\chi}\left (  { \kappa^{-2}}  -  \lambda\, \kappa^2 \right  ) \left (\hat{ \mathbf E} \cdot \hat{\mathbf{t}}_m \right )\right]<0\,.
\end{align}
We point out the following important features:
\begin{enumerate}
\item Here, unlike the isotropic case, the dispersion for $\mathbf E \propto [010]$ varies with the strength of $\gamma$. In addition to this, because of the breaking of inversion symmetry, a term linear in the external electric field is present, which leads to:
\begin{align}
\hat{\mathbf{E}} \cdot \hat{\mathbf{t}}_{1/3} = \sqrt{\frac{ 1}{3}}\, , \quad
\hat{\mathbf{E}} \cdot \hat{\mathbf{t}}_{2/4} =- \sqrt{\frac{ 1}{3}} \,.
\end{align} 
Hence, a transition from the $0$-flux phase to a $\pi$-flux phase occurs (absent in the isotropic limit). The instability develops along the  $\Gamma-Y$ high symmetry direction.
This is in contrast to the isotropic limit, which also shows the feature that the dispersion is independent of $\gamma$. 

\item 
{For ${\bf{ E}}\propto [011]$, 
the speed of the emergent photons vanishes along the $L-\Gamma$ high symmetry direction for a critical value
$ \gamma_c $. In this limit,
\begin{align}
& \hat{\mathbf{E}} \cdot \hat{\mathbf{t}}_{1} = \sqrt{\frac{ 2}{3}}\,, \quad
\hat{\mathbf{E}} \cdot \hat{\mathbf{t}}_{4} =- \sqrt{\frac{ 2}{3}}\,, \nn
& \hat{\mathbf{E}} \cdot \hat{\mathbf{t}}_{2/3}  = 0\,.
\end{align} 
For $\kappa=\lambda$, the critical value is $  \gamma_c^2 =1$ as $\mathcal{M}_{2,3} = 0$. For $\kappa > \lambda$ the critical value is $ \gamma_c^2  \geq 1 $ and for $\kappa < \lambda$, $ \gamma_c^2 < 1$.}

\item For ${\bf E}\propto [111]$, the critical value of $ \gamma_c^2 = 1.5$ for isotropic pyrochlores changes as $\kappa$ and $\lambda$ deviate from the value $1$. In fact, the value of $\gamma_c$ is dependent on the strength of $\kappa$ and $\lambda$. 

\end{enumerate}

Fig.~\eqref{fig:phasediag_noflux-flux} shows the phase diagram separating the $0$-flux and the selective $\pi$-flux for different directions of the electric field. Note that the flux phases are dependent on the directions of the electric field. Similar curves are obtained for different strengths of $\lambda$ (not shown here). 


\subsection{Spin Structure Factors}

\begin{figure*}
\includegraphics[width=0.3\linewidth]{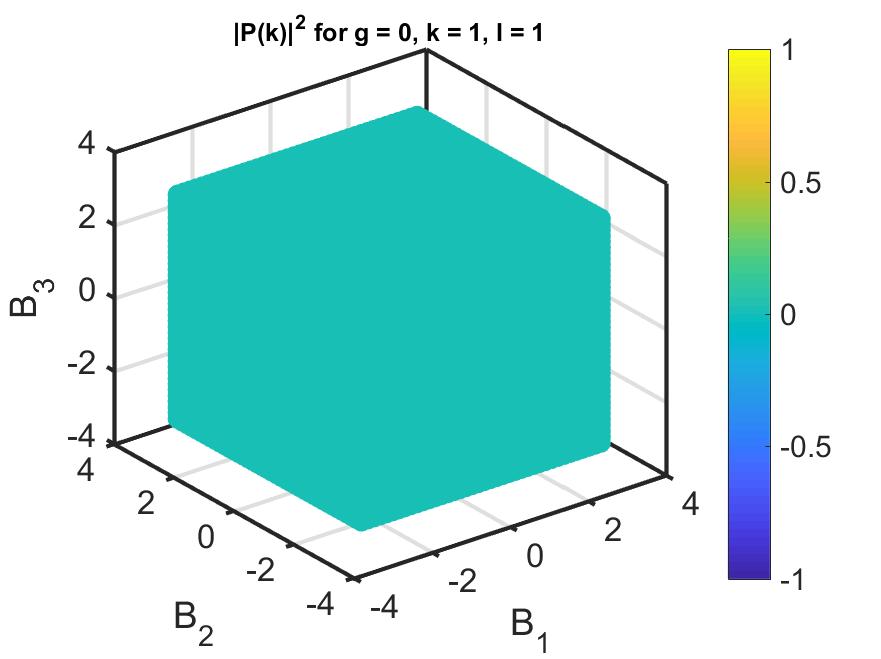}
\includegraphics[width=0.3\linewidth]{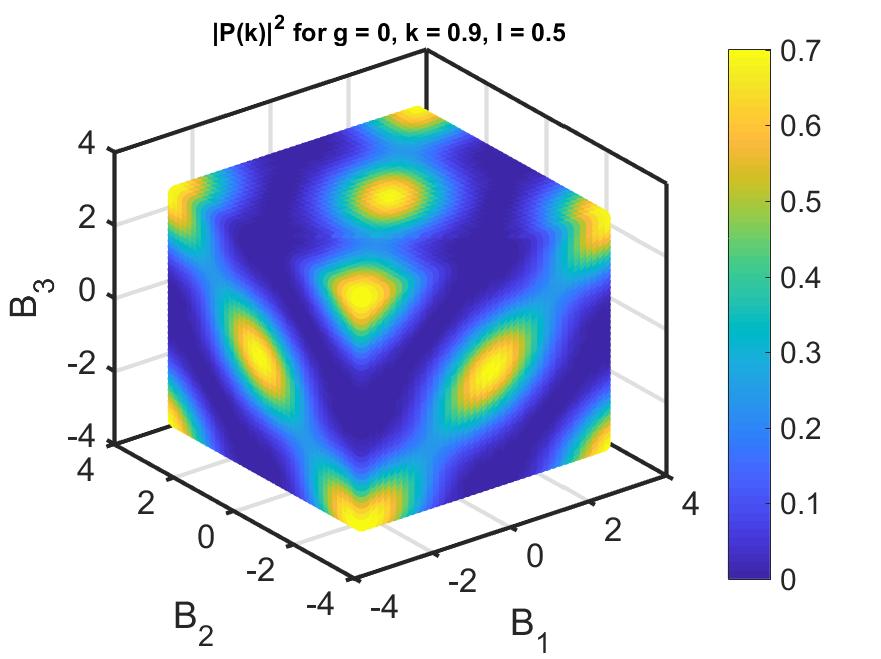}
\includegraphics[width=0.3\linewidth]{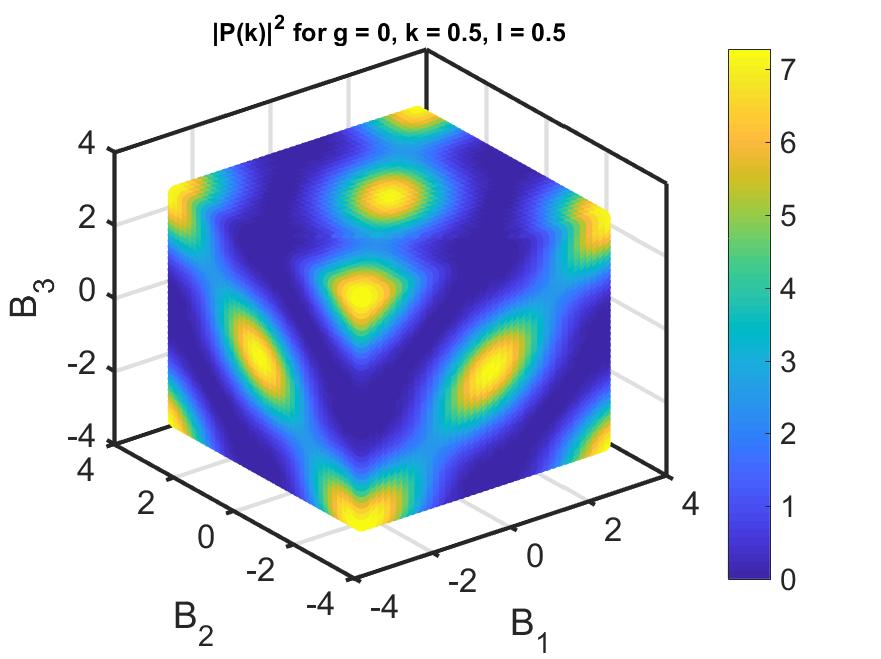}
\caption{The magnitude of the constant momentum-dependent polarization for $\mathbf E =0$, and (a) $\kappa = \lambda = 1$; (b) $\kappa = 0.9, \lambda = 0.5$; (c) $\kappa = \lambda = 0.5$.}
\label{fig:polkmag}
\end{figure*}
\begin{figure*}
\subfloat[]{
\includegraphics[width = 0.3\textwidth]{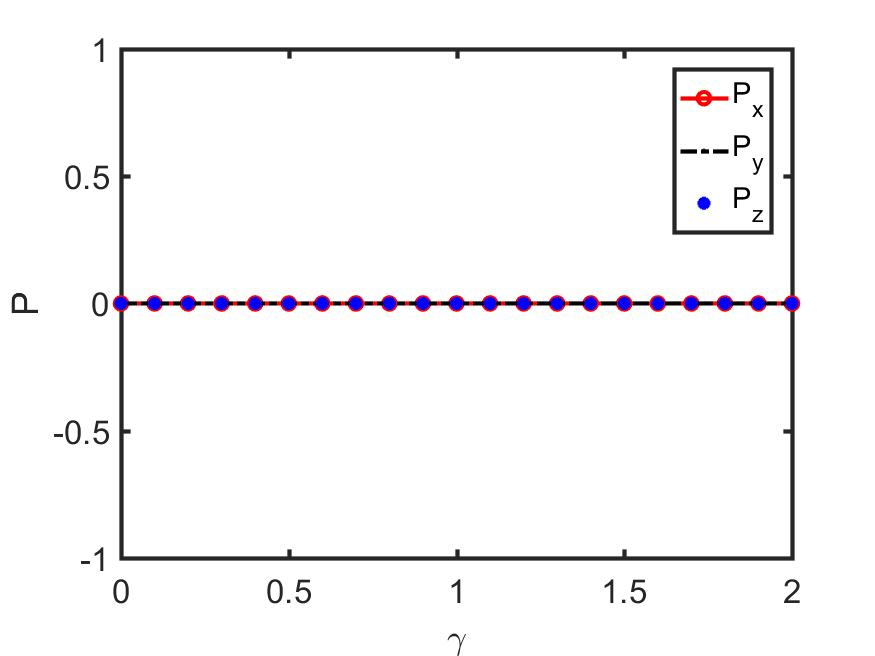}
} \quad
\subfloat[]{
\includegraphics[width = 0.3 \textwidth]{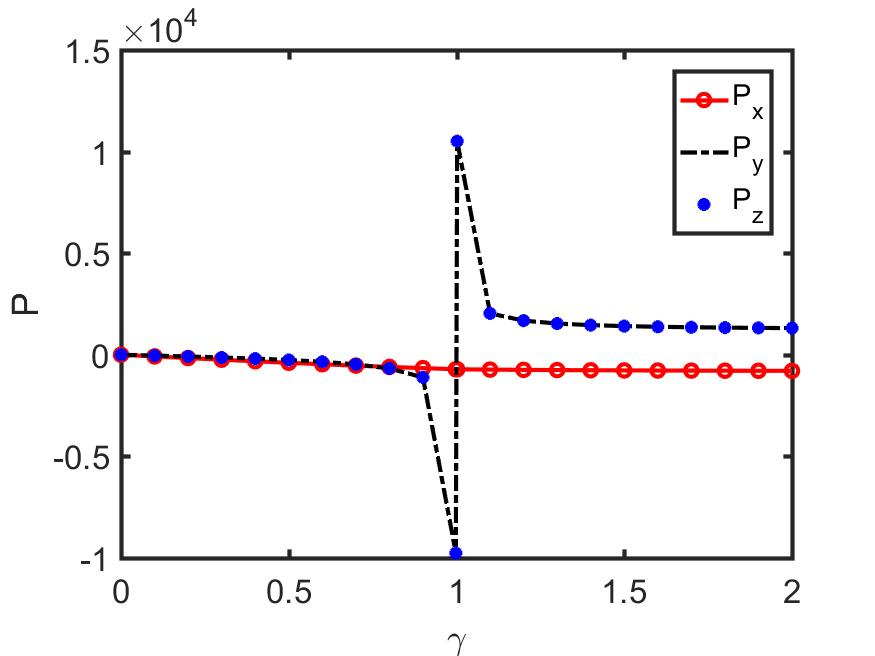}
} \quad
\subfloat[]{
\includegraphics[width = 0.3\textwidth]{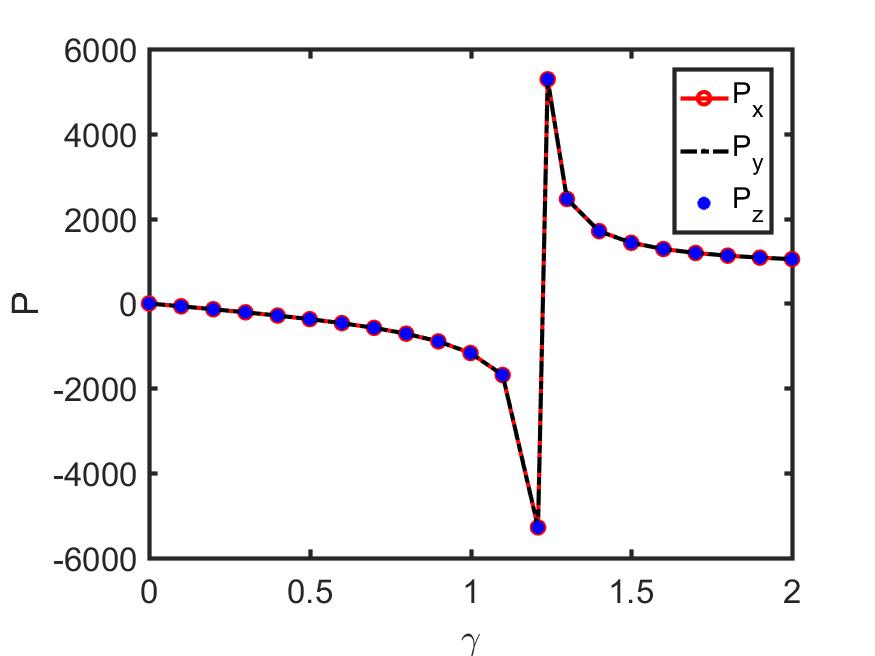}
}
\caption{The components of the polarization per site along the three directions of the electric field, for the isotropic pyrochlore case, with (a) $ \mathbf E \propto [010]$; (b) $\mathbf E \propto [011]$; (c) $\mathbf E  \propto [111]$. }
\label{fig:pol1}
\subfloat[]{
\includegraphics[width =  0.3\textwidth]{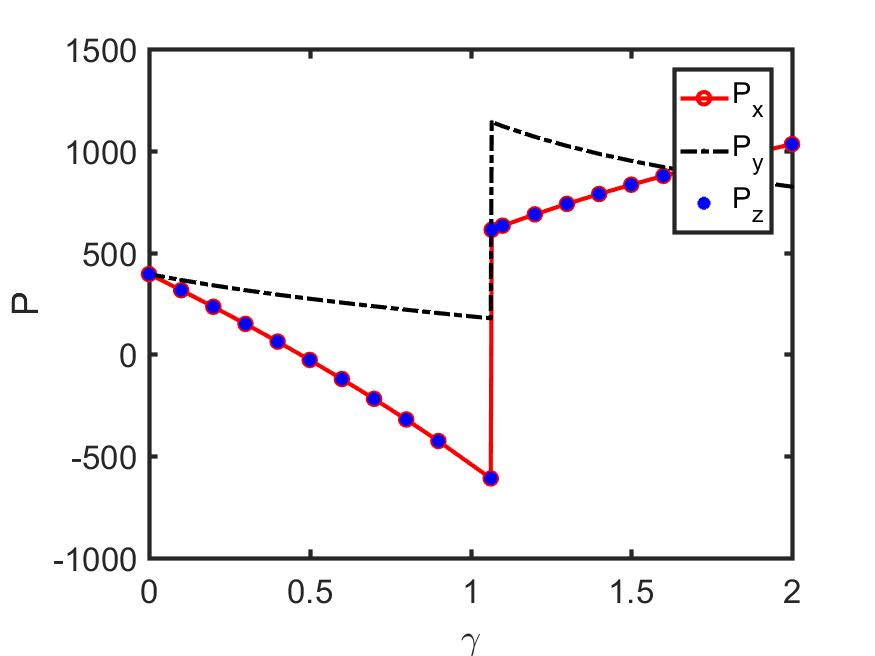}
} \quad
\subfloat[]{
\includegraphics[width =  0.3\textwidth]{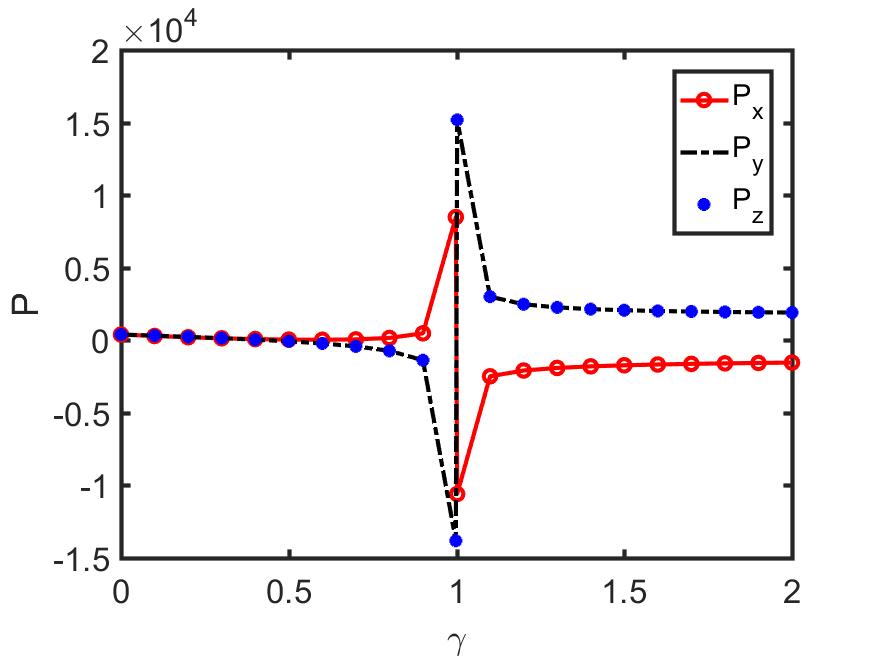}
} \quad
\subfloat[]{
\includegraphics[width =  0.3\textwidth]{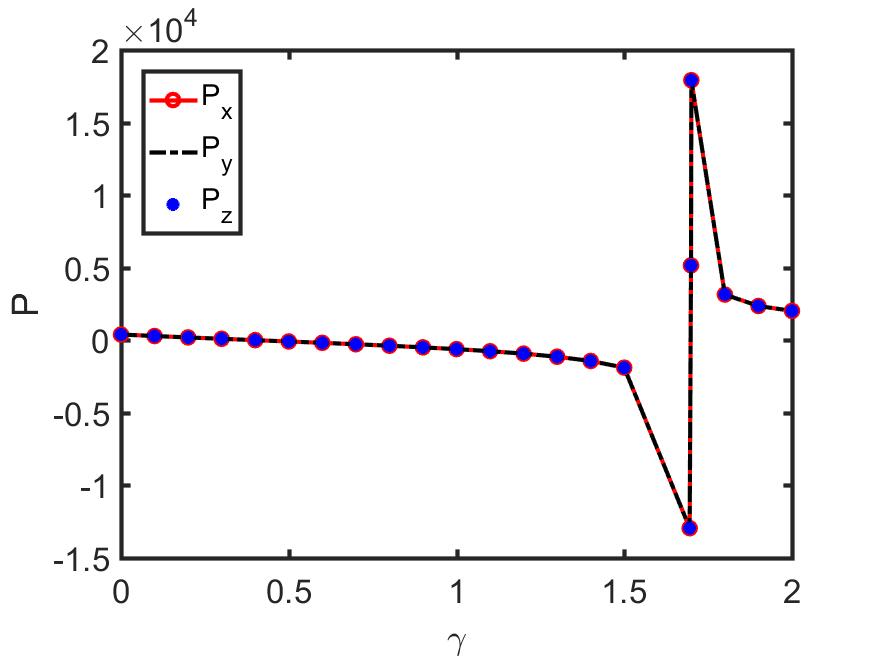}
}
\caption{The components of the polarization per site along three directions of the electric field, for the breathing pyrochlore case, with $\kappa = \lambda = 0.5$ and (a) $\mathbf E \propto  [010]$;(b) $\mathbf E \propto  [011]$; (c) $\mathbf E \propto  [111]$.}
\label{fig:pol2}
\end{figure*}

\begin{figure*}
\subfloat[]
{\includegraphics[width=0.35\linewidth]{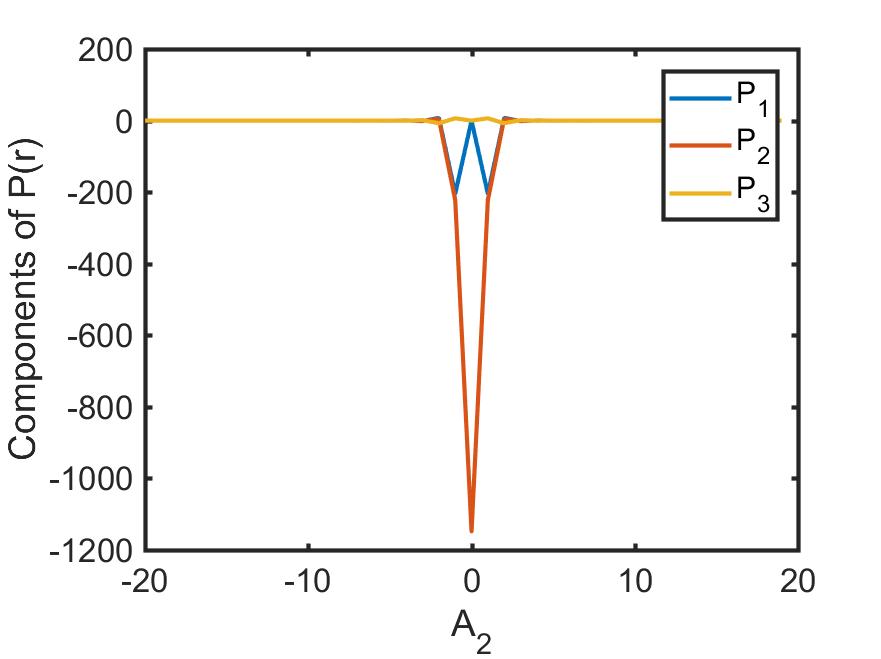}}
\quad
\subfloat[]
{\includegraphics[width=0.35\linewidth]{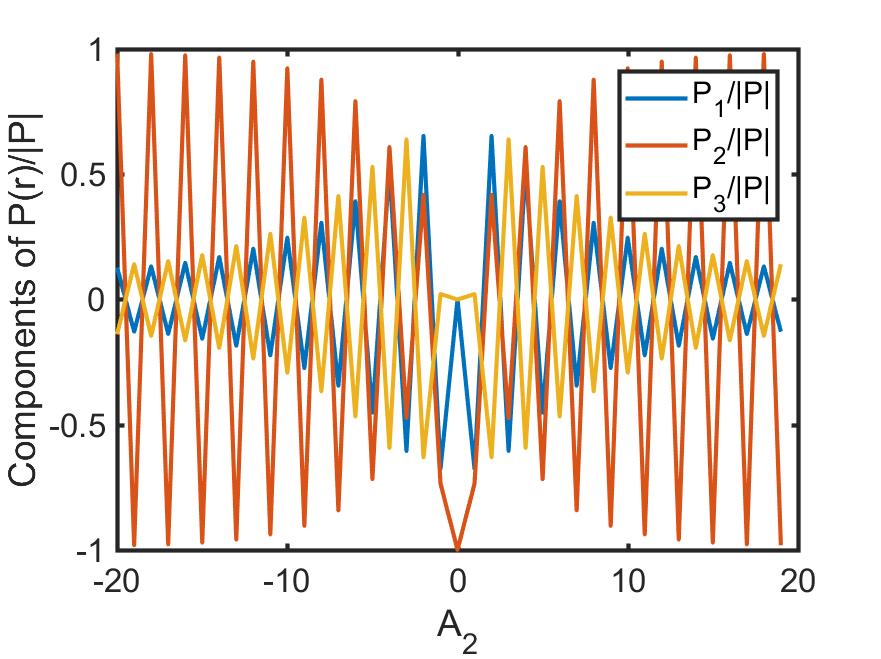}}
\\ \subfloat[]
{\includegraphics[width=0.35\linewidth]{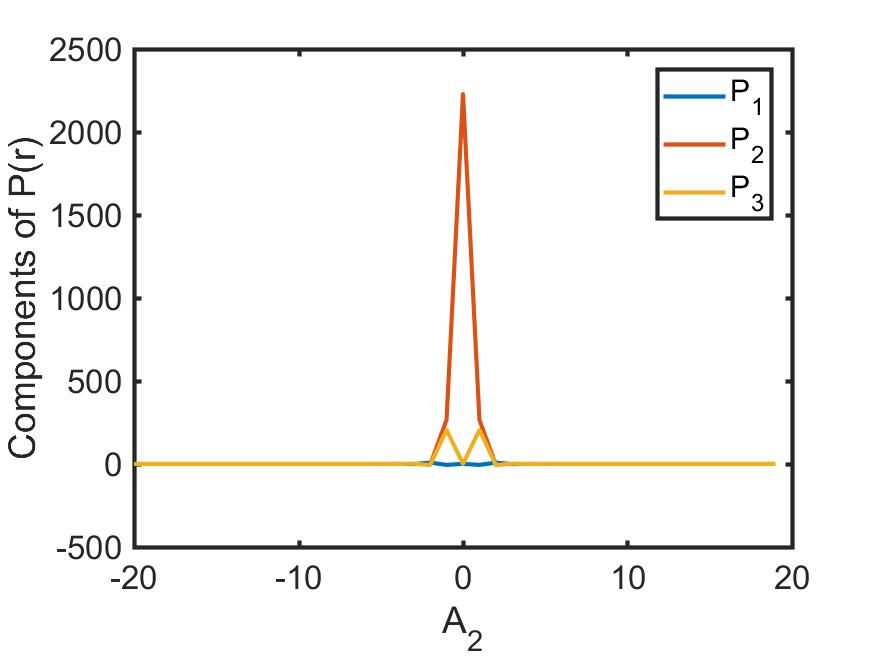}}
\quad \subfloat[]
{
\includegraphics[width=0.35\linewidth]{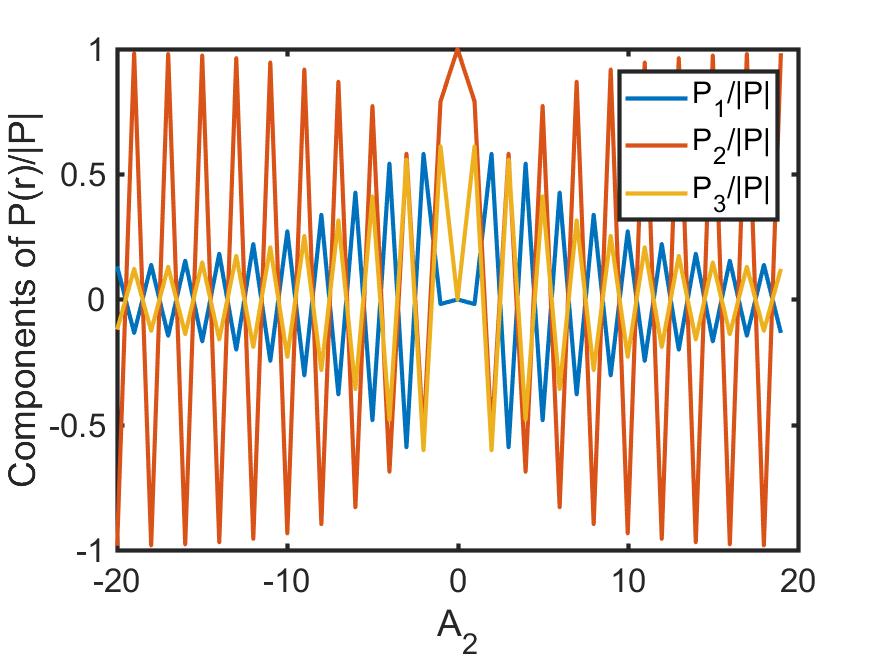}}
\caption{The components of polarization along $A_2$ lattice direction, for $\mathbf E \propto [010]$, $\kappa = \lambda = 0.5$ and fixed $A_1$ and $A_3$ values. Panels (a) and (b) correspond to $\gamma = 1.063$, panels (c) and (d) correspond to $\gamma = 1.066$.}
\label{fig:realpol_part}
\end{figure*}

The emergent photons can be detected in neutron scattering experiments, by measuring the equal-time (or energy-integrated) spin structure factors. Theoretically, this is given by:
\begin{align}
I^{\alpha \beta} (\mathbf{k}, t=0) =  \int d\omega~ I^{\alpha \beta} (\mathbf{k}, \omega)\,, 
\label{eq_et_sf}
\end{align}
where
\begin{align}
I^{\alpha \beta} (\mathbf{k}, \omega) =  \int dt~e^{-i\omega t} \left< s^\alpha(-\mathbf{k},t) s^\beta(\mathbf{k},0)\right>
\label{eq_dy_sf}
\end{align}
is the dynamic spin structure factor.
Defining the coordinate system:
$$ \mathbf{x} \parallel \mathbf{k} , \quad \mathbf{y} \parallel \bm{\eta}_\nu \times \mathbf{k} , \quad \mathbf{z} \parallel \bm{\eta}_\nu\, ,$$
where $\bm{\eta}_\nu \perp \mathbf{k}$ is the neutron polarization direction, we focus on the spin-flip channel characterized by $ \hat{\alpha} = \hat{\beta} = \hat{ \mathbf{y} }$. Performing similar computations as in Ref.~\onlinecite{etienne} (replacing their $\mathcal{M}_m $ with the defined in Eq.~\eqref{eq_Mmm}) lead to the final form:
\begin{align}
I^{yy} (\mathbf{k}, t=0)
& = \sum_{\phi,m,n,l,l'} \frac{\left( \hat{\mathbf{t}}_m \cdot \hat{\mathbf{y}} \right) \left( \hat{\mathbf{t}}_n \cdot \hat{\mathbf{y}} \right) } {8\, \omega_\phi (\mathbf{k})} \, \eta_{l \phi}(\mathbf{k}) \,\eta^*_{l' \phi}(\mathbf{k})
\nn
&\hspace{ 1.8 cm} \times  \sqrt{\mathcal{M}_l\,\mathcal{M}_{l'}} \,Z_{ml}(\mathbf{k})\,Z_{l'n}(\mathbf{k}) 
   \,.
\label{eq_structurefactors}
\end{align}
This is plotted in Fig.~\eqref{fig:struct_fac} for $ {\bf E}\propto [010]$ and $ {\bf E}\propto[011]$. 
In Fig.~\eqref{fig:struct_fac}(b), we have used $\kappa=\lambda$, for which $\gamma_c =1$. The structure factor is found to be symmetric along the $[\bar{1} 0 1]$ plane, which is not the case for the $[h 0 l]$ plane. 
It is worth mentioning here that the structure factors, with and without the $\pi$-flux, are the same. This is because in a U(1) gauge theory, the gauge bosons do not carry charge and hence cannot detect the presence of flux. On the other hand, the magnetic monopoles, which are the gapped excitations \cite{khomskii2012electric}, would be sensitive to such fluxes.

\subsection{Polarization Operator}


The polarization operator can be computed using ${{P}}_{eff}^\alpha = \frac{\partial }{\partial { E}_\alpha} \mathcal H_{eff}$, where $\alpha$ denotes the direction along which the component of the electric field $\mathbf E$ is considered. Alternatively, its expectation value can then be obtained by using Feynman-Hellmann theorem.

In the continuum limit, using Eq.~(\ref{cont-ham}), we have the expression:
\begin{align}
\label{pol-form}
& {{P}}_{eff} ^\alpha  
=\frac{\partial }{\partial { E}_\alpha} \sum_{{\bf{k}},\phi} \omega_\phi ({\bf{k}}) 
\left[ a^\dagger_{\phi \, {\bf k} } \,a_{\phi \, {\bf{k}} } + \frac{1}{2} \right] 
= \sum_{ {\bf k} } \tilde{ {P}}^\alpha( {\bf k} )\,,\nn
& \text{where }
\tilde {{P}}^\alpha( {\bf k} )  =  \sum_{\phi} \frac{\partial \omega_\phi ({\bf{k}})}{\partial { E}_\alpha}  
 \left[ a^\dagger_{\phi \, {\bf k} } \,a_{\phi \, {\bf{k}} } + \frac{1}{2} \right] .
\end{align}
Henceforth, we will drop the subscript ``${eff}$" from ${{\bf P}}_{eff}$ for simplicity of notations.
To obtain the expectation value, we note that $\left \langle \left [  a^\dagger_{\phi \, {\bf k} } \,a_{\phi \, {\bf{k}} }  + \frac{1}{2} \right] \right \rangle \approx 1$ as $T \rightarrow 0$.
We evaluate this numerically using the finite difference method. 

For the isotropic case, for $\mathbf E = 0$, ${\bf{P}}  =0$ for all momenta.  For the anisotropic case, a momentum-dependent polarization appears, an example of which is as shown in Fig.~\eqref{fig:polkmag}. The color represents the strength of the magnitude of the polarization vector at various points in the $3D$ momentum Brillouin zone. The magnitude depends on $\kappa$ and $\lambda$, as expected.

In Eq.~\eqref{eq_lattice_ham_matrixform}, as the $\mathcal{M}_m$'s change sign while we enter a $\pi$-flux phase (see the discussion in Sec.~\ref{piflux}), we can absorb the minus by using $- \cos \left( e_{\mathbf{s}, m} \right ) = \cos \left( e_{\mathbf{s}, m} - \pi \right ) $. The polarization vector can thus measure the electric part of the energy stored in the emergent electromagnetism. This also means that the expectation value of the classical part of the polarization is now non-zero. Hence, by tracking the behavior of the polarization as we change the parameters, we can identify this phase transition.

For $\mathbf{E} $ along $[010], \, [011]$ and $[111]$, the components of the net polarization per site are shown in Figs.~\eqref{fig:pol1} and~\eqref{fig:pol2}, as functions of $\gamma$. We observe the following cases:
\begin{enumerate}
\item
For a nonzero electric field along one of the crystallographic axes, e.g. $\mathbf{E}\propto [010]$, a polarization vector develops (unlike the isotropic case). The polarization components show sharp jumps at the transition from the $0$-flux phase to the $\pi$-flux phase.
\item For $\mathbf{E}\propto [011]$, the critical value at which the phase transition occurs is $ \gamma_c = 1$. We note that a $P_x$ component develops, which is roughly an order of magnitude lower than the $P_y=P_z$ component near the phase transition. For the isotropic case, this $P_x$ component does not undergo a discontinuity. For the breathing case, on the other hand, all the components show jump singularities.

\item
For $\mathbf{E}\propto [111]$, the polarization vector is directed opposite to the electric field in the $0$-flux phase, and at the phase transition it suddenly jumps to orient itself in the direction of the electric field in the $\pi$-flux phase. 
\end{enumerate}

To further understand how the polarization vector behaves (e.g. whether it complies with a ferroelectric behaviour), we have plotted in Fig.~\eqref{fig:realpol_part} the components of the polarization vector along the lattice direction $A_2$ (for fixed values of $A_1 $ and $A_3$), in the $0$-flux and the $\pi$-flux phases, close to the transition.
The components along the $A_1, \, A_2,\, A_3$ directions are denoted by $P_1,\,P_2,\,P_3$ respectively.
 We find that at distances covering two adjacent lattice sites, the vectors are almost antiparallel, indicating an antiferroelectric behavior.


\subsection{Polarization in the Trivial Paramagnetic Phases}

\begin{figure}[htb]
\subfloat[]{
\includegraphics[width=0.45 \textwidth]{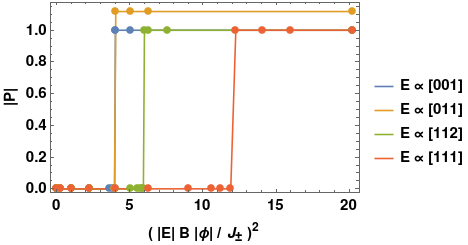}
}\\
\subfloat[]{
\includegraphics[width =0.17 \textwidth]{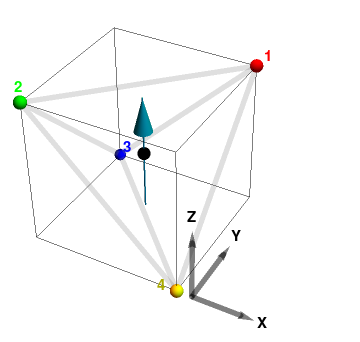}
}
\subfloat[]{
\includegraphics[width =0.17 \textwidth]{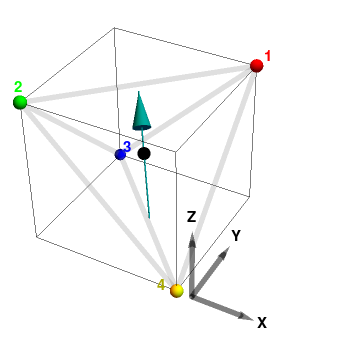}
}
\subfloat[]{
\includegraphics[width =0.17 \textwidth]{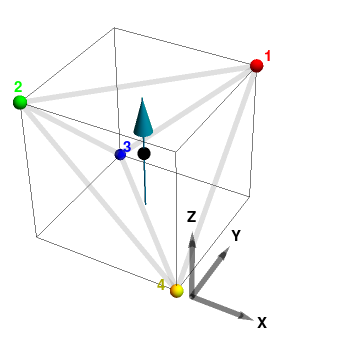}
}
\\
\subfloat[]{
\includegraphics[width =0.17 \textwidth]{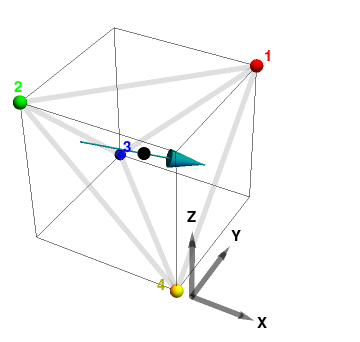}
}
\subfloat[]{
\includegraphics[width =0.17 \textwidth]{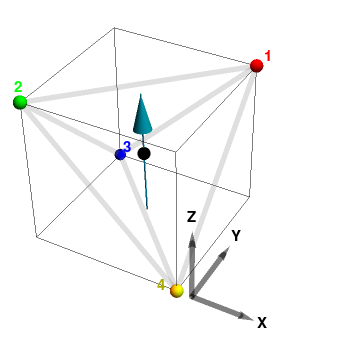}
}
\caption{The magnitude of the polarization for a single tetrahedron is shown in (a). The corresponding directions of polarization are shown in (b), (c), (d), (e), and (f), for the electric field along $[001],\,[011],\,[112],\,[111]$ and $[111]$ directions respectively. Note that (e) and (f) both correspond to $\mathbf E \propto [111]$, implying that the high field state is doubly degenerate for this case -- while the magnitudes of the polarization are the same, their directions are different.}
\label{fig_single}
\end{figure}

In a breathing pyrochlore, $\kappa=0$ or $\kappa=\infty$ denotes extreme cases with decoupled tetrahedra. Clearly as $\kappa$ is changed from the isotropic ($\kappa=1$) to either of the anisotropic limits, as shown in Fig.~\eqref{fig:phasediag_savary}, the system undergoes a transition from the U(1) QSL to a trivial paramagnet.  
In the extreme case of decoupled tetrahedra for $\kappa=0$, the up-tetrahedra are completely decoupled and the U(1) QSL is absent. One can ask about the dependence of the polarization on the external electric field for such a decoupled tetrahedron. We can compute the polarization very easily for these phases, as outlined in Appendix~\ref{appenpara}. The results are plotted in Fig.~\eqref{fig_single}. As a function of field strength, the ground state changes from one with zero polarization to one with finite polarization. This is signaled by a jump in the magnitude of the polarization as is evident from the first panel. The magnitude of the electric field required for the jump depends on the direction of the field.
The above trivial paramagnet is stable to small $\kappa$, and finally gives way to the U(1) QSL in presence of finite $J_\pm$.

\section{Conclusion}
\label{conclude}

In this paper, we have extended the study of the effects of
an externally applied uniform electric field $\mathbf E $ on the minimal quantum spin ice
Hamiltonian, exhibiting a U(1) QSL ground state, to the case of breathing pyrochlores.
We have computed the dispersion of the emergent photons, which shows birefringence, and have pointed out the differences from the isotropic limit. We have identified the transition points to $\pi$-flux phases as we increase the strength of the external electric field, which are the points where the photon velocity vanishes. Such phases are dependent of the direction of $\mathbf E$, and we found an emergent $\pi$-flux phase even for $\mathbf{E} \propto [010]$, which is nonexistent in the isotropic limit. We have showed the features expected to be observed in neutron scattering experiments, designed to detect the spin structure factors.
Finally, we have computed and elucidated the polarization behavior of the material. The contrasting features in the isotropic and anisotropic cases have been exhibited.

We have touched upon the gMFT formulation in Appendix~\ref{appenpara}, which can be used to chalk out the phase diagram in the presence of a nonzero $\mathbf{E}$, just like was done for the zero electric field case in Ref.~\onlinecite{savary2016}. We have applied it for the simple case of the trivial paramagnetic phase composed of decoupled tetrahedra. A more detailed computation for the possible $\pi$-flux phases is left for future work.

\section{Acknowledgments}

We  thank  Roderich Moessner and Subhro Bhattacharjee for suggesting this problem and collaborating in the initial stages of the project. We also thank P. V. Sriluckshmy for participating in the numerics.
We  are  grateful to \'Etienne Lantagne-Hurtubise and Jeffrey G. Rau for useful discussions.

\appendix

\section{Local Spin Quantization Axes}
\label{appen_latticedetails}

\begin{figure}[h]
\subfloat[]{
	\includegraphics[width = 0.25 \textwidth]{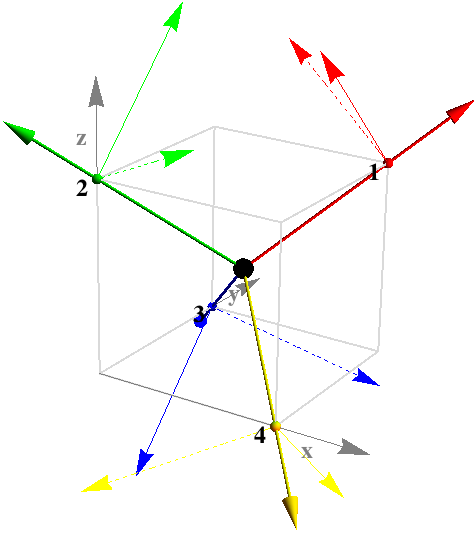}}
\subfloat[]{	
\includegraphics[width=0.25\textwidth]{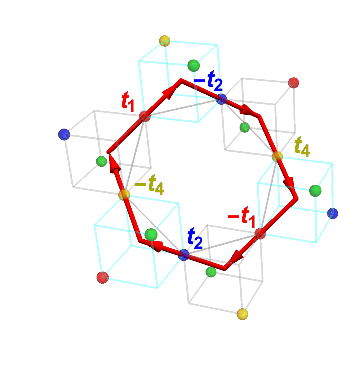}}
\caption{(a) The vectors depicting the local spin quantization axes for a tetrahedron of a breathing pyrochlore lattice. The thick lines denote the $\hat{\bf t}_m$'s, the thin lines denote the $\hat{\bf y}_m$'s, and the dashed lines denote the $\hat{\bf x}_m $'s. These directions are given by Eqs.~(\ref{eq_t_axes}), (\ref{eq_x_axes}) and (\ref{eq_y_axes}).
(b) Hexagon (of an isotropic pyrochlore) whose normal is $\hat{\bf t}_3$.}
\label{fig_1}
\end{figure}

The axes are denoted by $(\hat{\bf t}_m , \hat{\bf x}_m, \hat{\bf y}_m )$, are shown in Fig.~\eqref{fig_1}(a). These are explicitly given  by: 
\begin{align}
& \hat{\bf t}_1=\frac{1}{\sqrt{3}}[111]\,,\quad\hat{\bf t}_2=\frac{1}{\sqrt{3}}[\bar1\bar11]\,,\nonumber\\ ~
&\hat{\bf t}_3=\frac{1}{\sqrt{3}}[\bar11\bar1]\,, \quad\hat{\bf t}_4=\frac{1}{\sqrt{3}}[1\bar1\bar1]\,;
\label{eq_t_axes}
\end{align}
\begin{align}
&\hat{\bf x}_1=\frac{1}{\sqrt{6}}[\bar211]\,, \quad \hat{\bf x}_2=\frac{1}{\sqrt{6}}[2\bar11]\,,\quad\nonumber\\
&\hat{\bf x}_3=\frac{1}{\sqrt{6}}[21\bar1]\,, \quad \hat{\bf x}_4=\frac{1}{\sqrt{6}}[\bar2\bar1\bar1]\,;
\label{eq_x_axes}
\end{align}
\begin{align}
& \hat{\bf y}_1=\frac{1}{\sqrt{2}}[0\bar11]\,,\quad \hat{\bf y}_2=\frac{1}{\sqrt{2}}[011]\,,\quad\nonumber\\
& \hat{\bf y}_3=\frac{1}{\sqrt{2}}[0\bar1\bar1]\,,\quad  \hat{\bf y}_4=\frac{1}{\sqrt{2}}[01\bar1]\,.
\label{eq_y_axes}
\end{align}
We will denote the $a^{\text{th}}$ component of the three-component vector $ \hat{\mathbf{t}}_m $ by
$  \hat{\mathbf{t}}_m^a$. 

The FCC primitive lattice vectors are given by:
\begin{align}
\mathbf{A}_1 = \frac{a}{2}\,[101]\,, \quad \mathbf{A}_2 = \frac{a}{2}\,[110]
\,, \quad \mathbf{A}_3 = \frac{a}{2}\,[011]\,,
\label{eqfcc}
\end{align}
such that a real space lattice vector is $\mathbf{R} = \sum \limits _i n_i\, \mathbf{A}_i$. The basis vectors of the reciprocal FCC lattice are given as:
\begin{align}
\mathbf{B}_1 = \frac{2\pi\, [1 \bar 1 1]}{a} \,,
\quad \mathbf{B}_2 = \frac{2\pi\, [11 \bar 1]}{a} \,,
\quad \mathbf{B}_3 = \frac{2\pi\,  [\bar 111]}{a} \,,
\label{eqfcc2}
\end{align}
formed from $\mathbf{B}_1 =  \frac{2 \pi \left ( \mathbf{A}_2 \times  \mathbf{A}_3  \right )}
{\mathbf{A}_1 \cdot (\mathbf{A}_2 \times  \mathbf{A}_3)} $ and so on. Hence, the momentum vectors are given by:
\begin{align}
{\mathbf k} & = \sum_i q_i\, {\mathbf B}_i \,,
\end{align}
such that $-\frac{1}{2} < q_i < \frac{1}{2}$ forms the first Brillouin zone. 
We set $a = 1$ in our calculations. In Figs.~\eqref{fig:freq2} and~\eqref{fig:freq1}, the high symmetry points are represented in the original ${\bf k}_x, {\bf k}_y, {\bf k}_z$ basis.


There are four types of hexagons which can be labeled by the ``normal" vectors as follows:
\begin{align}
&\hat{\bf t}_1~{\rm involving}~[234]~~~\{\hat{\bf t}_2, \hat{\bf t}_3, \hat{\bf t}_4\}\,\nn
&\hat{\bf t}_2~{\rm involving}~[134]~~~\{\hat{\bf t}_1, \hat{\bf t}_2, \hat{\bf t}_4\}\,, \nn
&\hat{\bf t}_3~{\rm involving}~[124]~~~\{\hat{\bf t}_1, \hat{\bf t}_2, \hat{\bf t}_4\}\,,\nn
&\hat{\bf t}_4~{\rm involving}~[123]~~~\{\hat{\bf t}_1, \hat{\bf t}_2, \hat{\bf t}_3\}\,.
\end{align}
To avoid confusion, note that $\hat{\bf t}_m \cdot \hat{\bf t}_n  \neq 0$, and hence we call $\hat{\bf t}_n $ a ``normal" vector to label the corresponding hexagon in the sense that its sides are made up of $\hat{\bf t}_m$'s such that $m \neq n$. The hexagon centers of the direct pyrochlore lattice correspond to sites of the dual pyrochlore lattice. It is also important to note that the translation operators, required to act on the magnetic monopole to translate it around a given hexagon, only involve the translation vectors of $\hat{\bf t}_m $'s corresponding to the participating hexagons (with each $\hat{\bf t}_m $ occurring twice). This is shown in Fig.~\eqref{fig_1}(b) for a hexagon (of an isotropic pyrochlore) whose normal is $\hat{\bf t}_3$.

\section{Continuum Limit of the Effective Hamiltonian}
\label{continuum}
Here we analyze the continuum limit of the system, with the effective continuum Hamiltonian:
\begin{align}
\mathcal{H}_{eff}^{cont} & = \frac{1}{2} \int ~ d^3{\bf r}~\left  [ \mathcal{U} ~{\bf b}^2 + 
\sum \limits  _{\alpha, \beta =x,y,z}  e^\alpha ~\mathcal{K}^{\alpha\beta} ~e^\beta \right  ]\,,
\end{align}
where $\{e_x,e_y,e_z\}$ are the Cartesian components of the emergent electric field, $\mathcal{U} = \frac{4 \, U}{3\,l}$  ($l$ is a lattice length), and the matrix $\mathcal{K}$ is given by:
\begin{widetext}
\begin{align}
\mathcal{K} & = \frac{4\,g}{3\, l} \left( \begin{array}{ccc}
\mathcal{K}_1
&   \mathcal{K}_2\,  \hat{E}_x \,\hat{E}_y -  \mathcal{K}_3 \, \hat{E}_z &
 \mathcal{K}_2\, \hat{E}_x\, \hat{E}_z - \mathcal{K}_3\, \hat{E}_y\\
 \mathcal{K}_2\, \hat{E}_x \,\hat{E}_y
 - \mathcal{K}_3\, \hat{E}_z & \mathcal{K}_1 
& \mathcal{K}_2  \, \hat{E}_y \, \hat{E}_z - \mathcal{K}_3\, \hat{E}_x \\
 \mathcal{K}_2\, \hat{E}_x \hat{E}_z - 
 \mathcal{K}_3\, \hat{E}_y 
& \mathcal{K}_2\, \hat{E}_z \,\hat{E}_y -  \mathcal{K}_3\,  \hat{E}_x
 & \mathcal{K}_1
\end{array}\right) ,\nn
& \mathcal{K}_1= \kappa^3+\kappa^{-2} \,,
\quad   \mathcal{K}_2 =2 \, \gamma^2 \left (\lambda^2\kappa+\kappa^{-2} \right ),\quad 
 \mathcal{K}_3 =    \gamma \left (\kappa^{-2}-\lambda \, \kappa^2\right ).
\end{align}
\end{widetext}
The eigenvalues of this matrix are positive definite, and hence using the Cholesky decomposition \cite{golub2012matrix} of symmetric positive-definite matrices, we can write $\mathcal{K}=\Xi\cdot\Xi^T$,
where $\Xi$ is a lower triangular matrix. Finally, the photon dispersion relation is obtained by requiring that the terms which do not conserve photon number vanish. This is given by:
\begin{align}
	\omega_\phi(\mathbf{k}) =  |\xi_\phi (\hat{\mathbf{k}})|\, |\mathbf{k}|\,,
\end{align}
where $\xi_\phi^2(\hat{\mathbf{k}})$ denote the eigenvalues of the Hermitian, positive-definite matrix
\begin{align}
Q(\hat{\mathbf{k}}) = \mathcal{U}~ \Xi^T \begin{pmatrix}
1 - \hat{k}_x^2     & -\hat{k}_x \, \hat{k}_y & -\hat{k}_x\, \hat{k}_z \\
-\hat{k}_x \, \hat{k}_y & 1 - \hat{k}_y^2     & -\hat{k}_y\, \hat{k}_z \\
-\hat{k}_x \, \hat{k}_z & -\hat{k}_y \, \hat{k}_z & 1 - \hat{k}_z^2
\end{pmatrix} \Xi\,,
\end{align}
which only depends on the direction of $\mathbf{k}$. 

\section{Mean-field Computations in the Trivial Paramagnetic Phases}
\label{appenpara}

For the QSI Hamiltonian in Eq.~\eqref{breathing_haminfield}, we can apply the gauge mean field theory (gMFT) mapping as follows \cite{savary2012coulombic}:
\begin{align}
Q_{ \bf{ r}}=\eta_{ \bf{ r}}\sum \limits _{m=1}^4 s^z_{ \bf{ r}+\eta_r  \,{\mathbf{t}}_m } \,,
\quad { \mathbf{t}}_m=\frac{ \sqrt{ 3 }\,{\hat{\mathbf{t}}}_m } {4} \,, 
\label{icerules}
\end{align}
where $\eta_{ \bf{ r}}=\pm 1$ for the two sublattices of the diamond lattice
\footnote{The gMFT formulated by Savary \textit{et al}, in Ref.~\onlinecite{savary2012coulombic}, used the convention of choosing the local spin quantization axes with the vectors $ \hat{\bf e}_\mu \, \left [ \mu \in \left (0,1,2,3 \right ) \right ] $. They are related to our choice (see Eq.~\eqref{eq_t_axes}) by the following:
\begin{align}
 \hat{\bf e}_0 =  \hat{\bf t}_1\,,\quad   \hat{\bf e}_1 =  \hat{\bf t}_4\,,\quad 
 \hat{\bf e}_2 =  \hat{\bf t}_3 \,,\quad  \hat{\bf e}_3 =  \hat{\bf t}_2\,.
\end{align}
}. 
$Q_{ \bf{ r}}$ counts the number of magnetic monopoles. The spin flip operators get mapped to:
\begin{align}
& s^+_{ \bf{ r},\mathbf{r} + \mathbf{t}_m} =\Psi_{ \bf{ r}}^\dagger \, g_{ \bf{ r},\mathbf{r} + \mathbf{t}_m}\, \Psi_{ \mathbf{r} + \mathbf{t}_m}\,,\nonumber\\
& s^-_{ \mathbf{r} + \mathbf{t}_m} =\Psi_{ \mathbf{r} + \mathbf{t}_m}^\dagger\, g^\dagger_{ \bf{ r},r+\mathbf{t}_m}\, \Psi_{ \bf{ r}}\,,
\end{align}
where ${ \bf{ r}} \in$ up tetrahedron. The $U(1)$ gauge invariance implies, under gauge transformation, $\Psi$ and $g$ transforms as:
\begin{align}
\Psi_{\bf{ r}}\rightarrow	\Psi_{\bf{ r}}\, e^{-i\chi_{\bf{ r}}}\,,
~~~~~~~~~~g_{\bf{ r}r'}\rightarrow g_{\bf{ r}r'}\,e^{i(\chi_{\bf{ r}}-\chi_{\bf{ r}'})}\,.
\end{align}
Hence, the Hamiltonian for the breathing pyrochlore case, in absence of an electric field, becomes :
\begin{widetext}
\begin{align}
H&=\frac{J_{zz}}{2}\sum_{{\bf{ r}}~\in~u}
Q_{\bf{ r}}^2+\frac{\kappa J_{zz}}{2}\sum_{{\bf{ r}}~\in~d}
Q_{\bf{ r}}^2
-J_{\pm}\left[\sum_{{\bf{ r}}\in u}  \sum_{m\neq n}\Psi^\dagger_{\mathbf{r} + \mathbf{t}_n}\left(g^\dagger_{\bf{ r},\mathbf{r} + \mathbf{t}_n}g_{\bf{ r},\mathbf{r} + \mathbf{t}_m}\right)\Psi_{\mathbf{r} + \mathbf{t}_m}+\kappa\sum_{{\bf{ r}}\in d}\sum_{m \neq n}\Psi_{\bf{ r}- \mathbf{t}_n }^\dagger\left(g^\dagger_{\bf{ r},\mathbf{r} - \mathbf{t}_n} g_{\bf{ r},\mathbf{r} - \mathbf{t}_m}\right)\Psi_{\bf{ r}-  \mathbf{t}_m}\right],
\end{align}
\end{widetext}
where ${\mathbf{r} + \mathbf{t}_m}$ refers to the centre of a down-tetrahedron, when ${\bf{ r}}$ is the centre of an up-tetrahedron. Similarly, ${\bf{ r}- \mathbf{t}_m}$ refers to the centre of an up-tetrahedrpn when ${\bf{ r}}$ is the centre of a down-tetrahedron.
The first term may be thought as a chemical potential for the magnetic monopoles, which penalizes having such excitations. The second term makes such monopoles hop.

Within the gMFT approximation \cite{savary2012coulombic,savary2016,sungbin}, the Hamiltonian is decoupled into gauge and spinon sectors as follows:
\begin{align}
 \Psi^\dagger \Psi g^\dagger g & \approx \Psi^\dagger \, \Psi  \,  \langle g^\dagger \rangle \,  \langle g\rangle 
+ \langle \Psi^\dagger \,  \Psi\rangle \, g^\dagger  \, \langle g\rangle + \langle \Psi^\dagger\,  \Psi\rangle\,  \langle \, g^\dagger \rangle g\nonumber\\
& ~~~~~~~~~~~~~~~~~~~~~~~~-2\, \langle \Psi^\dagger \Psi\rangle\,  \langle\, g^\dagger \rangle  \, \langle g\rangle\,,
\nn s^zs^z   & \approx \langle s^z\rangle s^z \,.
\end{align}
Using this mapping, the transverse term of the polarization  operator of Eq.~(\ref{eq_ptrans}) for a single up tetrahedron can be written as:
\begin{widetext}
\begin{align}
{P}_x =&B\left[\Psi^\dagger_{\mathbf{0} + \mathbf{t}_4} 
\left (g^\dagger_{\bf 0,\mathbf{0} + \mathbf{t}_4} \ , g_{\bf 0,\mathbf{0} + \mathbf{t}_1}\right )\Psi_{\mathbf{0} + \mathbf{t}_1}-\Psi^\dagger_{\mathbf{0} + \mathbf{t}_3}
\left (g^\dagger_{\bf 0,\mathbf{0} + \mathbf{t}_3} \, g_{\bf 0,\mathbf{0} + \mathbf{t}_2} \right )\Psi_{\mathbf{0} + \mathbf{t}_2}\right]+{\rm h.c.}\,,\nn
{P}_y =&B\left[\Psi^\dagger_{\mathbf{0} + \mathbf{t}_3}
\left (g^\dagger_{\bf 0,\mathbf{0} + \mathbf{t}_3}\,  g_{\bf 0,\mathbf{0} + \mathbf{t}_1})\Psi_{\mathbf{0} + \mathbf{t}_1}-\Psi^\dagger_{\mathbf{0} + \mathbf{t}_4}(g^\dagger_{\bf 0,\mathbf{0} + \mathbf{t}_4}g_{\bf 0,\mathbf{0} + \mathbf{t}_2} \right )\Psi_{\mathbf{0} + \mathbf{t}_2}\right]+{\rm h.c.}\,,\nn
{P}_z = &B\left[\Psi^\dagger_{\mathbf{0} + \mathbf{t}_2}\left (g^\dagger_{\bf 0,\mathbf{0} + \mathbf{t}_2} \, g_{\bf 0,\mathbf{0} + \mathbf{t}_1}\right )\Psi_{\mathbf{0} + \mathbf{t}_1}-\Psi^\dagger_{\mathbf{0} + \mathbf{t}_4}  \left (g^\dagger_{\bf 0,\mathbf{0} + \mathbf{t}_4} \, g_{\bf 0,\mathbf{0} + \mathbf{t}_3} \right )\Psi_{\mathbf{0} + \mathbf{t}_3}\right]+{\rm h.c.}\,.
\label{pol-mft}
\end{align}
\end{widetext}
Note that both the monopoles are created on the neighboring down-tetrahedra as ${\bf 0}$ refers to the centre of the up-tetrahedron. Similar expressions can be written for the down tetrahedron which creates monopoles on the centres of the neighboring up-tetrahedra.


As we go across the transition to the paramagnetic phases, $\langle \Psi\rangle$ becomes nonzero selectively on either the up or down tetrahedra . Hence, within gMFT, $\langle \Psi\rangle$ is set to be a constant nonzero number $\phi$ (assuming $\langle \Psi\rangle$ condenses at zero momentum).
For the gauge fields, due to Higg's mechanism, it is fine to disregard their fluctuations. The mean field values of the amplitudes $\langle g\rangle$, depends on the projective realization of the symmetry and here we shall only discuss the effect of the translation symmetry, keeping in mind the possible rotation symmetry along the axis of the applied electric field.  

the polarisation operators of Eq.~(\ref{pol-mft}) effectively become:
\begin{align}
{P}_x =&B \,|\phi|^2\left[(g^\dagger_{\bf 0,\mathbf{0} + \mathbf{t}_4}g_{\bf 0,\mathbf{0} + \mathbf{t}_1})-(g^\dagger_{\bf 0,\mathbf{0} + \mathbf{t}_3}g_{\bf 0,\mathbf{0} + \mathbf{t}_2})\right]+{\rm h.c.}\,,\nn
{P}_y =&B  \, |\phi|^2\left[(g^\dagger_{\bf 0,\mathbf{0} + \mathbf{t}_3}g_{\bf 0,\mathbf{0} + \mathbf{t}_1})-(g^\dagger_{\bf 0,\mathbf{0} + \mathbf{t}_4}g_{\bf 0,\mathbf{0} + \mathbf{t}_2})\right]+{\rm h.c.} \,,\nn
{P}_z =&B  \, |\phi|^2\left[(g^\dagger_{\bf 0,\mathbf{0} + \mathbf{t}_2}g_{\bf 0,\mathbf{0} + \mathbf{t}_1})-(g^\dagger_{\bf 0,\mathbf{0} + \mathbf{t}_4}g_{\bf 0,\mathbf{0} + \mathbf{t}_3})\right]+{\rm h.c.}\,.
\label{pol-PM}
\end{align}

To understand the effect of the electric field in a paramagnetic phase, it is sufficient to consider the effect of polarization on a single tetrahedron, as it consists of either up or down decoupled tetrahedra. Here we consider the effect of electric field in the paramagnetic phase consisting of decoupled up-tetrahedra. Using Eq.~(\ref{pol-PM}), we get:
\begin{widetext}
\begin{align}
\mathcal{H}  
  =   & \frac{J_{zz}}{2}\, Q_{\bf 0 }^2
 -J_{\pm}  \, |\phi|^2\left[ g^\dagger_{ \bf 0 ,\mathbf{0} + \mathbf{t}_1 }\,g_{\bf 0 ,\mathbf{0} + \mathbf{t}_2} 
+  g^\dagger_{ \bf 0 ,\mathbf{0} + \mathbf{t}_2 }\,g_{\bf 0 ,\mathbf{0} + \mathbf{t}_3 } 
+  g^\dagger_{ \bf 0 ,\mathbf{0} + \mathbf{t}_3 }\,g_{\bf 0 ,\mathbf{0} + \mathbf{t}_4 } 
  \right]
 +  E_x\, {P}_x 
+ +  E_y\, {P}_y 
+  E_z\, {P}_z  \,.
\end{align}
\end{widetext}


\bibliography{biblio}
\end{document}